\documentclass[preprint]{elsarticle}
\usepackage{amsmath,epsfig,epstopdf, amssymb}
\usepackage{amsthm} 
\usepackage{float}
\usepackage{xspace}
\usepackage{algorithmic}
\floatstyle{boxed}
\newfloat{algorithm}{htbp}{loa}
\floatname{algorithm}{Algorithm}
\floatstyle{boxed}

\usepackage{wrapfig}
\newcommand{\rimp}{\Rightarrow}
\usepackage{multirow}

\newtheorem{thm}{Theorem} [section]
\newtheorem{lem}[thm]{Lemma}

\newtheorem{cor}{Corollary}

\newtheorem{defn}{Definition}

\title {Finding Minimum and Maximum Termination Time of Timed Automata Models with Cyclic Behaviour}
\author  {Omar Al-Bataineh$^1$, Mark Reynolds$^2$, and Tim French$^2$ 
\\ $^1$Nanyang Technological University, Singapore \\  $^2$University of Western Australia, Australia} 
\date{}

\begin{document}

\label{firstpage}


\begin{abstract}
The paper presents a novel algorithm for computing worst case execution time (WCET) or maximum termination time of real-time systems using the timed automata (TA) model checking technology. The algorithm can work on any arbitrary diagonal-free TA and can handle more cases than previously existing algorithms for WCET computation, as it can handle cycles in TA and decide whether they lead to an infinite WCET. We show soundness of the proposed algorithm and study its complexity. To our knowledge, this is the first model checking algorithm that addresses comprehensively the WCET problem of systems with cyclic behaviour. In \cite{Behrmann2001} Behrmann et al provide an algorithm for computing the minimum cost/time of reaching a goal state in priced timed automata (PTA).  The algorithm has been implemented in the well-known model checking tool UPPAAL to compute the minimum time for termination of an automaton. However, we show that in certain circumstances, when infinite cycles exist, the algorithm implemented in UPPAAL may not terminate, and we provide examples which UPPAAL fails to verify.

\end{abstract}

\maketitle

\section{Introduction}

In this paper, we consider the problem of computing the ``worst case execution
time" (WCET) in timed automata. Given a timed automaton $\mathcal{A}$ with a start location $l_s$ and a final location $l_f$, this problem asks to compute an upper bound on the time
needed to reach the final location $l_f$ from the start location $l_s$. 
The problem is easy to solve in the case of acyclic TA \cite{AlBataineh14}, but cycles might introduce an unbounded WCET, that needs to be detected on-the-fly during the analysis. 
In general, WCET analysis is undecidable: it is undecidable to determine whether or not an execution of a system will eventually halt. However, for TA models one can use model-checking techniques to analyse the system and compute the WCET. 

Typically, the infinite state-space of a timed transition system is converted into an equivalent
finite state-space of a symbolic transition system called a zone graph \cite{Dill1990,MCBook}.
In a zone graph, zones (i.e. sets of valuations of the timed automaton clocks) are used to denote symbolic states. 
The zone graph has been successfully used for the verification of safety and liveness properties of timed automata. Although the zone graph is precise enough to preserve the reachability properties in TA, it is too abstract to infer continuous time progress.
At each step of the successor computation, the generated zones
are extrapolated (abstracted) using a set of extrapolation operators
and then canonicalized (tightened) in order to obtain a unique representation of the resulting zones.
A test for inclusion of zones is then applied to check whether the new generated zone at a particular control location in the graph is already covered by some previously generated zones associated with that location. This helps to ensure termination of the analysis of TA even when infinite cycles exist.

\begin{figure} 
  \begin{minipage}[b]{0.4\linewidth}
    \centering
    \includegraphics[width= 2in]{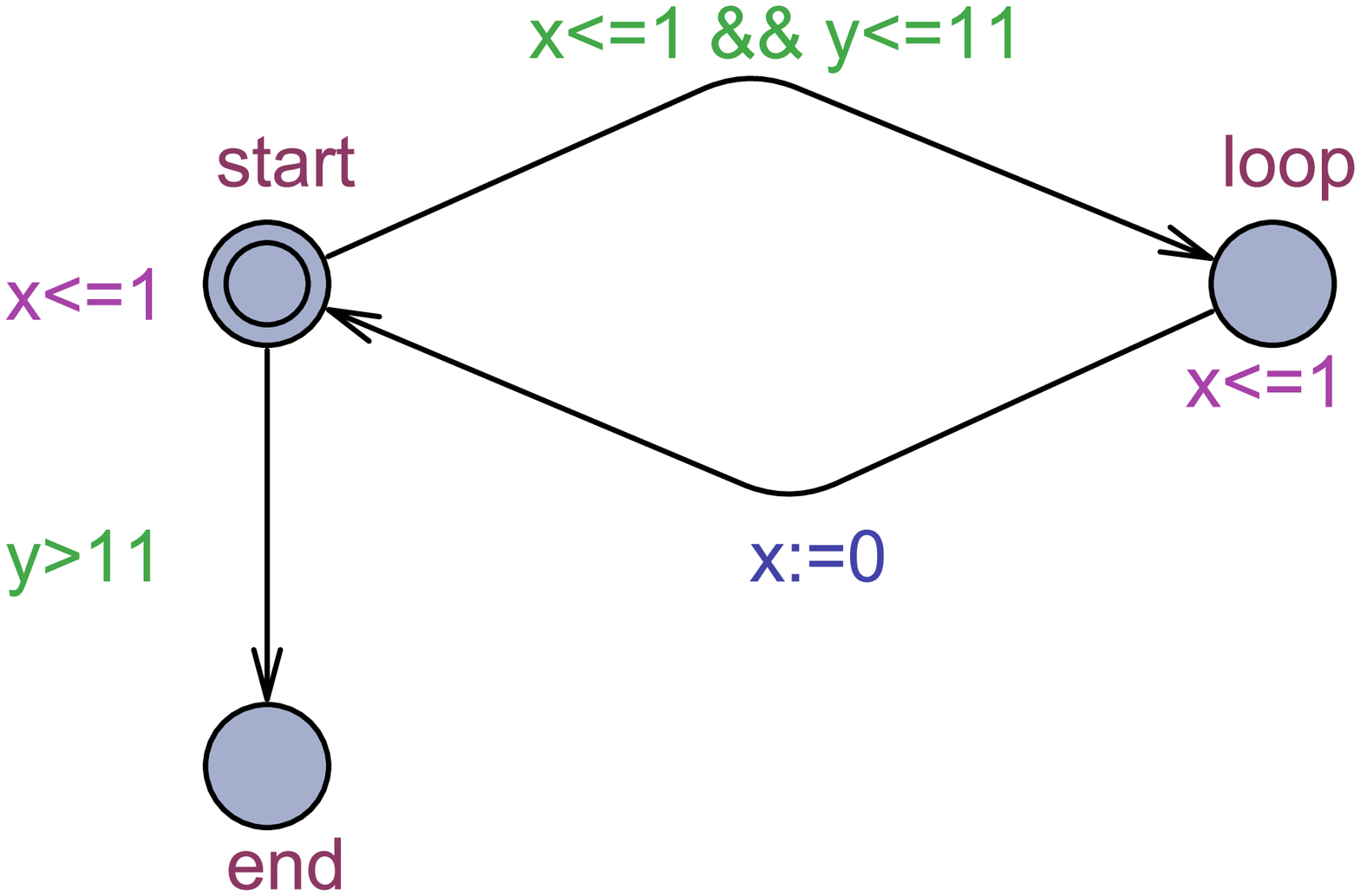}
    \caption{$\mathcal{A}_1$: an automaton with finite cycle}
    \label{fig:finiteone}
  \end{minipage}
  \hspace{1.3cm}
  \begin{minipage}[b]{0.4\linewidth}
    \centering
    \includegraphics[width= 2in]{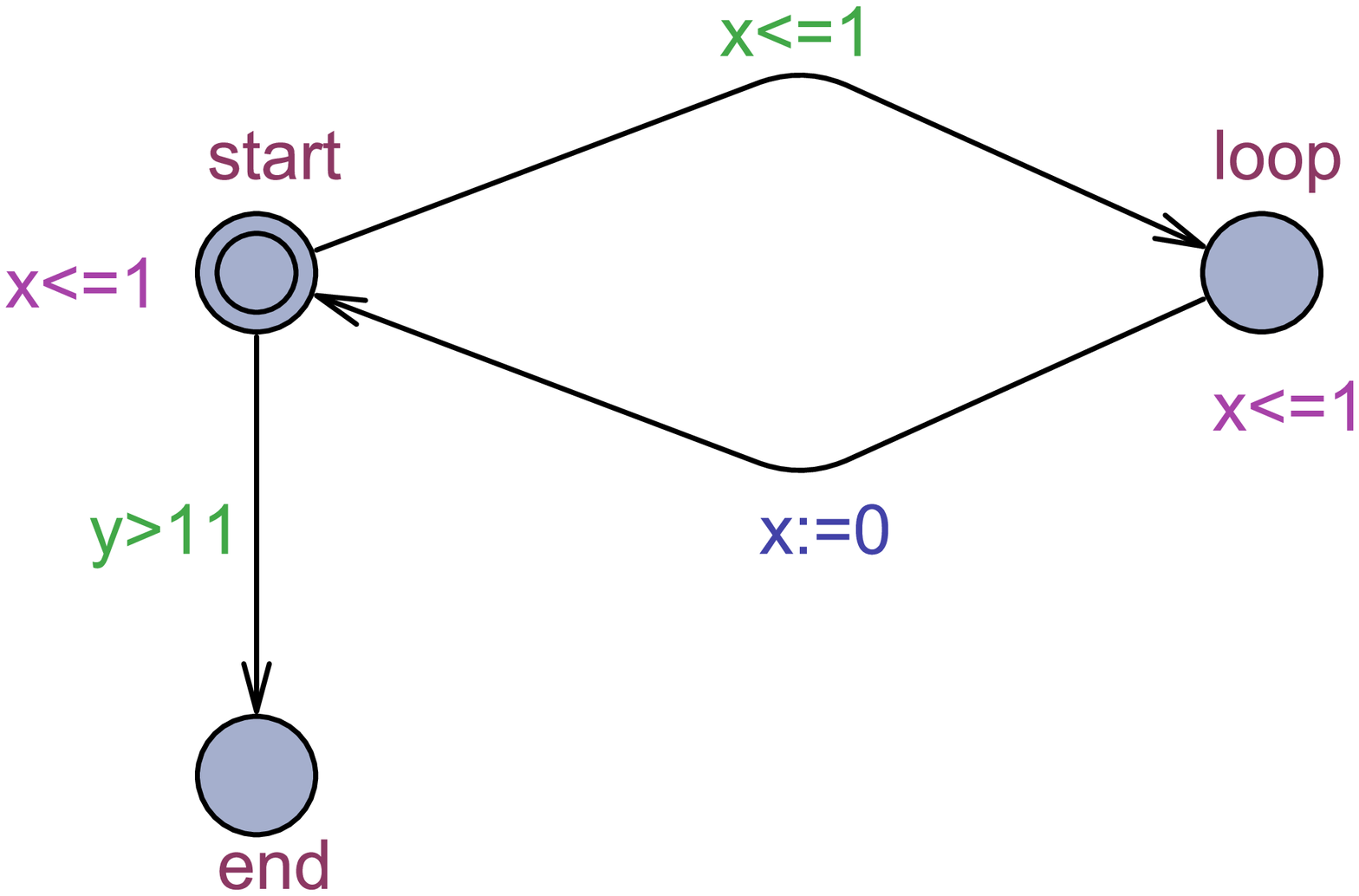}
    \caption{$\mathcal{A}_2$: an automaton with infinite cycle}
    \label{fig:infinitefirst}
  \end{minipage}
\end{figure}

However, the classical abstraction used for verification of reachability problem \cite{Bengtsson04} is not correct for WCET and BCET computation, as they give abstract zones and hence result in abstract values of the execution times.
To demonstrate the problem, we give in Figures \ref{fig:finiteone} and \ref{fig:infinitefirst} two automata where both generate identical zone graphs when applying the standard zone approach for reachability analysis.
The automaton $\mathcal{A}_1$ represents an automaton with finite cycle  
where $\mathit{WCET}(\mathcal{A}_1) = 12$. 
For this automaton, the standard zone approach can compute correctly the WCET without involving any extra check. On the other hand, the automaton $\mathcal{A}_2$ represents an automaton with an infinite cycle where $\mathit{WCET}(\mathcal{A}_2) = \infty$. 
For this automaton, the zone approach for reachability analysis fails to give the correct answer for WCET since it returns 12 instead of $\infty$. Note that if we disable extrapolation during the analysis, the search may not stop and we may not be able to obtain an answer.

In a previous work \cite{AlBataineh14}, we proposed a zone-based solution to the problem of computing WCET of real-time systems modelled as TA. The proposed solution allows one to compute the WCET of TA in only one run of the zone construction instead of making repeated guesses (guided by binary search) and multiple model checking queries as done in \cite{Alex04}. However, in \cite{AlBataineh14} we limit applicability of our solution to timed automata without infinite cycles.
In the present paper, we give a more general solution to the problem that can work on any arbitrary diagonal-free TA\footnote{A class of TA in which the test of the form $x- y \sim c$ is disallowed, where $x, y$ are clock variables, $c$ is a constant, and $\sim \in \{<, \leq, =, >, \geq \}$.} including those containing infinite cycles. Infinite cycles indeed make the computation of WCET difficult because zone
extrapolation techniques are necessary to compute a finite state-space, and extrapolation prevents
a straightforward computation of the WCET.
The main contribution of the paper is therefore to
propose an extrapolation technique that is compatible with the WCET computation. 
More precisely, we give the special conditions needed to define a forward zone-based reachability algorithm that terminates and computes the correct maximal time. Thus, the provided solution can be a significant break-through in computing WCET. The proposed extrapolation technique is an interesting addition to the collection of techniques for TA analysis. It is particularly useful because it improves zone extrapolation, that is one of the weak points of TA symbolic analysis. 

In \cite{Behrmann2001} Behrmann et al propose an algorithm that aims to provide a solution to the minimum cost/time reachability problem in Uniformally Priced Timed Automata (UPTA) in the presence of extrapolation. The algorithm has been implemented in the well-known model checking tool UPPAAL to compute the minimum time for termination of an automaton. However, the extrapolation step is not detailed in \cite{Behrmann2001} and the implementation in UPPAAL is often unable to terminate when the model has some cycles. The key difficulty in developing a solution to the minimum/maximum termination time problems using the zone approach is to define an abstraction of zones that guarantees termination of the algorithm, while keeping information precise for the extra clock that is used to compute the execution time of the automaton. This involves adapting two classical operations on zones: extrapolation and canonicalization. The later was forgotten in \cite{Behrmann2001} leading to non termination. We give a number examples by which we demonstrate how and why existing algorithms for computing BCET and WCET fail (including the one being now used in the tool UPPAAL). 

\paragraph{Related Work}

It is claimed in \cite{Rein04} that model checking is inadequate for
WCET analysis. However,  in \cite{Alex04} Metzner showed that model checking can be used efficiently for WCET analysis. He used model checking to improve WCET analyses for hardware with caching. 
The use of timed automata (TA) and the model-checker UPPAAL for computing WCET on pipelined
processors with caches was reported in \cite{And10} where the METAMOC method is described. METAMOC
consists in: 1) computing the CFG of a program, 2) composing this CFG with a (network of timed automata)
model of the processor and the caches. Computing the WCET is then reduced to computing the longest path
(timewise) in the network of TA.

The work in \cite{Behrmann2001} uses a variant of timed automata
called  ``Priced Timed Automata'' and the DBM data structure to 
compute the minimum cost of reaching a goal state in the model.  
A priced timed automaton can associate costs with locations, where the costs are multiplied by the amount of time spent in a location.
An automaton may be designed so that the total cost corresponds to the execution time, and thus this approach may be used to calculate the best case execution time problem. However, the WCET problem is different than the BCET problem and needs special treatment during the analysis
in particular when there are cycles in the behaviour of TA.

In \cite{BehrmannLR05}  Behrmann et al. provide zone-based algorithms
for parameter synthesis for two  strict forms of TCTL properties: (1) $ \textbf{AF}_{\leq p} ~\phi$ and (2) $ \textbf{AG} (\psi \rimp \textbf{AF}_{\leq p} ~\phi)$.
The first form can be used to calculate the WCET of the given TA model.
However, none of these two forms can be used to directly calculate optimum time or BCET of the model.
The algorithms require the user to have some prior knowledge about the behaviour of the given model in the sense that the user has to identify the set of goal states (e.g. final states) in order to use a TCTL formula for calculating WCET. Moreover, it is not clear to us how this approach can
be used to handle TA with infinite cycles and whether it can detect the cases
where the WCET is infinity.

In \cite{AlBataineh14} Al-Bataineh et al present a solution to the problems of computing the shortest and the longest time taken by a run of a timed automaton from an initial state to a final state.  
The solution is conceptually a marked improvement over some earlier work on the problems \cite{Alex04}, in which repeated guesses (guided by binary search) and multiple model checking queries were effectively but inelegantly and less efficiently used; while in \cite{AlBataineh14} only one run of the zone construction is sufficient to yield the answers. However, the authors of \cite{AlBataineh14} limit applicability of their approach to timed automata without infinite runs.


The efficient verification of WCET of timed automata models with cyclic behaviour requires to detect on-the-fly the existence of infinite zeno runs (i.e. runs in which time cannot diverge) and infinite non-zeno runs (i.e. runs in which time can diverge) in the behaviour of the automaton under analysis. 
This is necessary in order to guarantee termination of the analysis.
Detection of infinite non-zeno runs
was already addressed in \cite{Alur94}. Their approach works on the region graph, but for correctness reasons, 
it cannot be used on (abstract) zone graphs. The trick involving 
adding an extra clock for non-zenoness is discussed in   \cite{Tripakis99verifyingprogress,Tripakis05checkingtimed,Alur04decisionproblems}. 
The problem of checking existence of zeno runs   
was formulated as early as in \cite{Tripakis99verifyingprogress}. 
A bulk of the literature for this problem also directs
to \cite{Gomez07,Courcoubetis1992,Rinast2012}. All of these solutions provide a sufficient-only condition for the absence
of zeno runs. However, the purpose of our work is to present the special conditions needed to define a forward zone-based reachability algorithm that terminates and computes the correct maximal time while using the abstract zone graph, which requires to handle on-the-fly infinite zeno runs and infinite non-zeno runs.

The structure of the paper is as follows. We begin in Section 2 by introducing the syntax and the semantics of TA and the syntax and the semantics of the zone graph. We then review the existing extrapolation procedures of TA and discuss their role in forward reachability algorithms. 
In Section 3, we discuss some interesting issues about the minimum cost reachability algorithm proposed by Behrmann et al \cite{Behrmann2001} and its implementation in UPPAAL. In Section 4, we introduce what we call partial extrapolation procedure of zones and prove its correctness. We also discuss cycles (loops) in TA and describe what we call fixed point abstraction to detect (on-the-fly) infinite cycles. In Section 5, we describe a model checking algorithm for computing WCET for the class of diagonal-free TA. In Section 6, we study the complexity of the algorithm. In Section 7, we describe an implementation of the algorithm using the model checker opaal and describe the associated verification results on a set of examples.  Finally, in Section 8, we draw some conclusions and discuss future directions.

\section{Preliminaries}

\subsection{Timed Automata}


Timed automata are an extension
of the classical finite state automata with clock variables to model timing
aspects \cite{Alur94}. 
Let $X$ be a set of clock variables, the clock valuation $v$ for the set $X$ is a mapping from
$X$ to $\mathbb{R}^{+}$ where $\mathbb{R}^{+}$ denotes the set 
of non-negative real numbers. 
 
\begin{defn}
A timed automaton $\mathcal{A}$ is a tuple $(\Sigma, L, L_{0}, L_{F}, X, I, E)$, where

\begin{itemize}

\item $\Sigma$ is a finite set of actions.

\item $L$ is a finite set of locations.

\item $L_{0} \subseteq L $ is a finite set of initial or starting locations.

\item $L_{F} \subseteq L $ is a finite set of final locations.

\item $X$ is a finite set of clocks.

\item $I: L \rightarrow \mathcal{C}(X)$ is a mapping from locations to clock constraints, called
the location invariant.

\item $E \subseteq L \times L \times \Sigma \times 2^{X} \times \mathcal{C}(X)$
is a finite set of transitions.
An edge $(l, l^{'}, a, \lambda, \phi)$ represents
a transition from location $l$ to location $l^{'}$ after performing action $a$.
The set $\lambda \subseteq X$ gives the clocks to be reset with this transition,
and $\phi$ is a clock constraint over $X$.   The clock constraint $\phi$ can be of the form:
$
\phi ::= x \prec c  \mid \phi_{1} \land \phi_{2}
$, where $x \in X$, $c \in \mathbb{N}$, and $\prec \in \{<, \leq, =, >, \geq \}$.
\end{itemize}
\end{defn}

We define the semantics of a timed automaton by an infinite labelled transition system.
The states in this system are tuples $(l, v)$, where $l$ is the current location of the 
automaton, and $v$ is a function that maps the clocks of the automaton to a non-negative
real number. 
The initial states are of the form $ (l_{0}, v_{0})$ 
where $l_{0}$ in $L_{0}$ and the valuation $ v_{0}(x) = 0$ for all $x \in X$.
With each transition we associate a clock constraint called a guard
and with each location we associate a clock constraint called its invariant.

Transitions of an automaton may include clock resets and guards which give conditions on the 
interval in which a transition can be executed.

\begin{defn}
Transitions in timed automata are of two forms:
\begin{enumerate}
\item delay transitions that model the elapse of time while staying at some location:
for a state $(l, v)$ and a real-valued time increment $\delta \geq 0$, $(l, v)  \xrightarrow{\delta} (l, v + \delta)$
if for all $v'$ with $v \leq v' \leq v+ \delta$, the invariant $I(l)$ holds.
\item action transitions that execute an edge of the automata: for a state $(l, v)$
and a transition $(l, l^{'}, a, \lambda, \phi)$ such that $v \models \phi$, 
$(l, v)  \xrightarrow{a} (l^{'}, v[\lambda:=0])$. 
\end{enumerate}

So for an automaton to move from a location to another a delay transition
followed by an action transition must be performed. We write this as $\xrightarrow{d_{i}} \xrightarrow{a_{i}}$. 
\end{defn}

\begin{defn}
A run of a timed automaton with an initial state $ (l_{0}, v_{0})$
over a timed trace $ \zeta= (t_{1}, a_{1}), (t_{2},a_{2}),...$ 
is a sequence of transitions of the form.
$$\langle l_{0}, v_{0} \rangle \xrightarrow{d1} \xrightarrow{a1} \langle l_{1}, 
v_{1}\rangle \xrightarrow{d2} \xrightarrow{a2} \langle l_{2}, 
 v_{2} \rangle,...$$
satisfying the condition $t_{i} = t_{i-1}+d_{i}$ for all $i \geq 1$ and that $l_{0} \in L_{0}$.
\end{defn}

Since the locations of an automaton are decorated with a delay-quantity and that transitions between locations are instantaneous, the delay of a run is simply the sum of the delays spent in the visited locations. 

\begin{defn}
Let $ r = \langle l_{0}, v_{0} \rangle \xrightarrow{d1} \xrightarrow{a1} \langle l_{1}, 
v_{1}\rangle \xrightarrow{d2} \xrightarrow{a2} \langle l_{2}, 
 v_{2} \rangle,...$ be a run in the set of runs $\mathcal{R}$. 
The delay of $r$, $delay(r)$, is the sum $\sum_{i=1}^{n} d_{i}$, where $n$ can be infinity. 
Hence, the problem of computing the BCET and WCET of $\mathcal{A}$ can be formalized as follows.
$$
\mathit{BCET} (\mathcal{A}) = \inf_{\forall {r \in \mathcal{R}^f}} (delay(r))
$$
$$
\mathit{WCET} (\mathcal{A}) = \sup_{\forall {r \in \mathcal{R}}} (delay(r)) 
$$  
\end{defn}

where $\mathcal{R}^f$ is the set of runs in $\mathcal{R}$ that reach a final location.
Of course, a valid WCET bound is $[0, \infty]$, and WCET can be infinity if there is an infinite non-zeno run (an infinite run in which time can diverge) \cite{Bowman2006,RodolfoPhD}. This  can happen if there is a reachable cycle that can be repeated infinitely often and that time can elapse between iterations.

\begin{defn} 
A cycle in a timed automaton is a finite sequence of edges where the source location of the first edge in the sequence is the target location of the last edge in the sequence. Let $\mathcal{A} = (\Sigma, L, L_{0}, L_{F}, X, I, E)$ be a timed automaton and let $m$ be a natural number such that $m \geq 1$.  We say that a sequence $(e_0, e_1,..., e_{m-1}) \in E^m$ is a cycle if $trg(e_i) = src(e_{i+1})$ for all $0 \leq i < m-1$ and $trg(e_{m-1}) = src(e_{0})$.
\end{defn}

\subsection{The Clock Zones and The Difference Bound Matrices}

The infinite state-space of a TA can be converted into an equivalent
finite state-space of a symbolic transition system called a zone graph \cite{Dill1990,MCBook}.
A state in a zone graph is a pair $(l, Z)$, where $l$ is a location in the TA model and $Z$ is a clock zone that represents a set of clock valuations at $l$.
Formally a clock zone is a conjunction of inequalities that compare either a clock value or the difference between two clock values to
an integer. In order to have a unified form for clock zones we introduce a reference clock $x_{0}$ 
to the set of clocks $X$ in the analyzed model that is always zero.
The general form of a clock zone can be described by the following formula.
$$
(x_{0} = 0) \land \bigwedge_{0\leq i \neq j \leq n} ( (x_{i}-x_{j}) \prec c_{i,j})
$$
where $x_{i}, x_{j} \in X$, $c_{i,j}$ bounds the difference between them, and $ \prec \in \{ \leq, <\}$.  
Consider a timed automaton $\mathcal{A} = (\Sigma, L, L_{0}, L_{F}, X, I, E)$, with a transition $e = (l, l^{'}, a, \lambda, \phi)$ in $E$.
We can construct an abstract zone graph $\mathcal{Z(A)}$ such that states of $\mathcal{Z(A)}$
are zones of $\mathcal{A}$. 
The clock zone $succ(Z, e)$ denotes the set of clock valuations $Z^{'}$ for which
the state $(l^{'}, Z^{'})$ can be reached from the state $(l, Z)$
by letting time elapse and by executing the transition $e$.
The pair $(l^{'}, succ(Z,e))$ represents the set of successors of $(l, Z)$
after firing the transition $e$ (see Section \ref{sec:ZoneAppraoch} for how to compute the suceesor of a zone $Z$ w.r.t a transition $e$). 

The most important property of zones is that they can be represented as matrices. Several algorithms
based on the notion of zones are implemented using the difference bound matrices (DBMs), which is the most commonly used data structure for the representation of zones.

A DBM is a two-dimensional matrix that records the difference upper bounds between clock pairs up to a
certain constant. In order to have a unified form for clock constraints in DBM matrix we introduce a reference clock $x_{0}$ with the
constant value 0. 
The element in matrix $D$ is of the form  $(d_{i, j}, \prec)$
where $x_{i}, x_{j} \in X$, $d_{i, j}$ bounds the difference between $x_i - x_j$, and $ \prec \in \{ \leq, <\}$.  
Each row in the matrix represents the bound difference between the value of the clock $x_{i}$ and all the other clocks in the zone,
thus a zone can be represented by at most $|X|^2$ atomic constraints. 
Since the variable $x_{0}$ is always 0, it can be used to express constraints
that only involve a single variable. 
For example, the element $(d_{i,0},\prec)$, means that we have the constraint $x_{i} \prec d_{i}$.
However, to obtain a unique representation of the matrix (zone) so that each atomic constraint in the matrix is in the tightest or canonical form,
most model checking tools for timed automata use the Floyd-Warshall algorithm \cite{Floyd1962}.
In fact, canonical forms simplify some operations over DBMs like the test for inclusion between zones.

\subsection{The Extrapolation Abstraction} \label{sec:extrapolation}

In the definition of timed automata, we allow clocks in the invariant
of a location and in the guards of the transitions to have arbitrary non-negative
real numbers, which makes the model checking problem of timed
automata seem intractable since the number of states is infinite.
To obtain a finite zone graph most model checkers use some kind of extrapolation of
zones. In the last two decades, there has been a considerable development in the extrapolation
procedure for TA for the purpose of providing coarser abstractions of TA \cite{RokickiPhD,Bengtsson04,Bouyer2004,Behrmann2006}. We review these procedures in the following subsections.

\subsubsection{Classical Maximal Bounds.}
One of the first proposed extrapolation algorithms for TA is the so-called $M$-extrapolation  \cite{Daws1998}, i.e. the zone is extrapolated with respect to the maximum constant
each clock is compared to in the automaton.
That is, if the clock is never compared to a
constant greater than $M$ in a guard or invariant, then the value of the clock will have no impact on the
computation of the automaton once it exceeds $M$. 
The $M$-extrapolation algorithm has been implemented in the early version of UPPAAL \cite{Beh04}.
The procedure to obtain the $M$-extrapolation of a given zone is to remove all upper bounds higher than the maximum constant 
and lowering all lower bounds higher than the maximum constant down to the maximum constant.

\begin{defn} \label{m-norm}
Let $Z$ be a zone represented by a DBM in a canonical form $D = (m_{i,j}, \prec_{i,j})_{i,j=0,..n}$ and $M$ be the largest integer constant that appears in the guard and the location
invariants of $\mathcal{A}$.  We can define the extrapolation function $\verb+Extra+_{M}(D^{'})$ of the zone $D^{'} = (m^{'}_{i,j}, \prec^{'}_{i,j})_{i,j=0,..n}$ as follows:
$$
(m^{'}_{i,j}; \prec^{'}_{i,j}) = 
\begin{cases}
 (\infty, <) &  \textrm{if $m_{i,j}> M$}, \\
  (-M, <) & \textrm{ if $m_{i,j} <-M$}, \\
  (m_{i,j}, \prec_{i,j})  & \textrm{otherwise.} \\
\end{cases}
$$
\end{defn}

\begin{lem} \cite{PetterssonPhD}
For diagonal-free TA, 
the symbolic set $(l, Extra_{M}(Z))$ and the transitions $\leadsto_{M}$ resulting from
the $M$-extrapolation are sound and complete with respect to reachability and the transition relation is finite.
\end{lem}

A maximal constant can be computed for each clock in the automaton in a similar way,
which could make the state space much smaller.
A considerable gain in efficiency can be obtained by analysing the graph of the automaton
and calculating maximum bounds specific for each clock and state of the automaton \cite{Behrmann2003}.
That is, the maximum constants not only depend of the particular clock but also of the particular
location of the TA. An even more efficient approach is the so called $LU$-extrapolation that distinguishes 
between upper and lower bounds \cite{Behrmann2006}. This is the method used in the current implementation of UPPAAL.

\subsubsection{Lower and Upper Maximal Bounds.} 

In \cite{Behrmann2006} it has been observed that by distinguishing the maximal lower and upper bounds to which clocks of the timed automaton are compared one can obtain a significantly coarser abstraction of TA.

\begin{defn} \label{LUNorm}
Let $Z$ be a zone represented by a DBM in a canonical form $D = (m_{i,j}, \prec_{i,j})_{i,j=0,..n}$. 
For each clock $x_{i} \in X$ in $\mathcal{A}$, the maximal lower bound $L(x_{i})$, (resp.  maximal upper bound of $x_{i}$  $U(x_{i})$))
is the maximal constant $M$ such that there exists a constraint $x>M$ or $x\geq M$ (resp. $x<M$ or $x \leq M$)
in a guard of a transition or in an invariant of some location in $\mathcal{A}$. If such a constant does not exist,
we set $L(x_{i})$, (resp. $U(x_{i}))$ to $-\infty$.
The $LU$-extrapolation of the zone $D^{'} = (m^{'}_{i,j}, \prec^{'}_{i,j})_{i,j=0,..n}$ can be defined as follows.
$$
(m^{'}_{i,j}; \prec^{'}_{i,j}) = 
\begin{cases}
 \infty &  \textrm{if $m_{i,j}> L(x_{i})$}, \\
  (-U(x_{j}), <) & \textrm{ if $-m_{i,j} > U(x_{i})$}, \\
  (m_{i,j}, \prec_{i,j})  & \textrm{otherwise.} \\
\end{cases}
$$
\end{defn}

Note that the $LU$-extrapolation benefit from the properties of the two different maximal bounds.
It does generalise the $M$-extrapolation (i.e. $\forall_{x \in X} (M(x) = \max (L(x), U(x)))$). 
For every zone $Z$, it holds that $Z \subseteq Extra_{M} (Z) \subseteq Extra_{LU} (Z)$ \cite{Behrmann2006}.
The experiments given in \cite{Behrmann2006} demonstrate the significant 
speedup obtained from using lower and upper bounds of clocks in the abstraction.
Note that the $M$-extrapolation and the $LU$-extrapolation operations will not preserve the canonical form of the DBM,
and in this case the best way to put the result back on canonical form is to use the Floyd-Warshall algorithm.

\begin{lem} \cite{Behrmann2006}
For diagonal-free TA the $LU$-extrapolation is sound, complete, finite with respect to reachability and effectively computable.
\end{lem}

However, in \cite{Behrmann2006} the authors have discussed also two other extrapolation procedures that can 
provide coarser abstraction of TA, namely $\verb+Extra+_{M}^{+} (Z)$ and $\verb+Extra+_{LU}^{+} (Z)$.
The improvement proposed in these procedures is based on the observation that when the whole zone is above the maximum bound of some clock, then one can remove some of the diagonal constraints of the zones, even if they are not themselves above the maximal bound. Formally, we can define the $\verb+Extra+_{M}^{+} (Z)$ operation as follows.
$$\label{MNorm+}
(m^{'}_{i,j}; \prec^{'}_{i,j}) = 
\begin{cases}
 \infty &  \textrm{if $m_{i,j}> M(x_{i})$}, \\
 \infty &  \textrm{if $-m_{0,i}> M(x_{i})$}, \\
  \infty &  \textrm{if $-m_{0,j}> M(x_{j})$}, i \neq 0 \\
  (-M(x_{j}), <) & \textrm{ if $-m_{i,j} > M(x_{j})$}, i =0 \\
  (m_{i,j}, \prec_{i,j})  & \textrm{otherwise.} \\
\end{cases}
$$
Similarly, we can define the  $\verb+Extra+_{LU}^{+} (Z)$ operation as follows.
$$ \label{LUNorm+}
(m^{'}_{i,j}; \prec^{'}_{i,j}) = 
\begin{cases}
 \infty &  \textrm{if $m_{i,j}> L(x_{i})$}, \\
 \infty &  \textrm{if $-m_{0,i}> L(x_{i})$}, \\
  \infty &  \textrm{if $-m_{0,j}> U(x_{j})$}, i \neq 0 \\
  (-U(x_{j}), <) & \textrm{ if $-m_{0,j} > U(x_{j})$}, i =0 \\
  (m_{i,j}, \prec_{i,j})  & \textrm{otherwise.} \\
\end{cases}
$$

\subsection{The Standard Zone-based Approach} \label{sec:ZoneAppraoch}

Before presenting our proposed solution to  the WCET problem it is necessary first to summarise
how the zone or DBM based successor computation can be performed. 
Let $D$ be a DBM in canonical form. We want to compute the successor
of $D$ w.r.t to a transition $e = (l, l^{'}, a, \lambda, \phi)$, let us denote it as $succ(D, e)$. The clock zone $succ(D, e)$  can be obtained using a number of elementary DBM operations which can be described as follows.

\begin{enumerate}

\item Let an arbitrary amount of time elapse on all clocks in $D$.
In a DBM this means all elements $D_{i, 0}$ are set to $\infty$.
We will use the operator $\Uparrow$ to denote the time elapse operation.

\item Take the intersection with the invariant of location $l$ to find the set of possible clock assignments
	  that still satisfy the invariant.

\item Take the intersection with the guard $\phi$ to find the clock assignments that are accepted
      by the transition.

\item  Canonicalize the resulting DBM and check the consistency of the matrix.

\item Set all the clocks in $\lambda$ that are reset by the transition to 0.


\item Take the intersection with the location invariant of the target location $l^{'}$.

\item Canonicalize the resulting DBM.

\item Extrapolate and canonicalize the resulting zone at the target location $l^{'}$ and check the consistency of the matrix.

\end{enumerate}

Combining all of the above steps into one formula, we obtain
$$
 \begin{array}[t]{l}
succ(D, e) = (\verb+Canon+(\verb+Extra+(\verb+Canon+((\verb+Canon+(((D ^{\Uparrow}) \land   I(l)) \land  \phi) [\lambda:=0]) \land  I(l^{'})))))
\end{array}
$$
where \verb+Extra+ represents an extrapolation function that takes as input a DBM and returns the $M$-form of the matrix, while \verb+Canon+ represents a canonicalization function that takes as input a DBM and returns a canonicalized matrix in the sense that
each atomic constraint in the matrix is in the tightest form,
$I(l)$ is the invariant at location $l$, and $\Uparrow$ denotes the elapse of time operation. 
Note that intersection does not preserve canonical form \cite{Bengtsson04}, so we should 
canonicalize $(((D ^{\Uparrow}) \land   I(l)) \land  \phi)$ before resetting any clock (if any). 
Since after executing the transition $e$ all the clocks in the automaton have to advance at the same rate.
After applying the guard, the matrix must be checked for consistency.
Checking the consistency of a DBM is done by computing the canonical form and then checking the diagonal for negative entries.
The resulting zone at step 5 needs to be intersected with  the clock invariant at the target location $l^{'}$ and extrapolating/canonicalization afterwards.  This is necessary in order to ensure that the guard $\phi$ and the reset operation $([\lambda:=0])$ implies the invariant at the target location.

Before proceeding further let us review first the three elementary operations that are used to construct the zone graph of a given automaton, which are the \textit{intersection} operation, 
the \textit{reset} operation, and the \textit{delay} operation or the \textit{elapse of time} operation.

\begin{defn}
(The intersection operation). 
We define $D = D^1 \land D^2$. Let $D^{1}_{i, j} = (c_1, \prec_1)$ and $D^{2}_{i, j} = (c_2, \prec_2)$. Then
$
D_{i,j} = (\min (c_1,c_2), \prec)$ where $\prec$ is defined as follows.
$$
\prec = 
\begin{cases}
   \prec_1 &  \textrm{if}~ c_1 < c_2, \\
   \prec_2 &  \textrm{if}~  c_2 < c_1, \\
   \prec_1 &  \textrm{if}~ c_1 = c_2 ~ \land \prec_1 = \prec_2, \\
   < &  \textrm{if}~  c_1 = c_2 ~ \land \prec_1 \neq \prec_2, 
\end{cases}
$$
\end {defn}

As mentioned before intersection does not preserve canonical form and the best way to put  the matrix back on canonical form is to use the Floyd-Warshall algorithm. However, the work in \cite{Zhao2005} presents an algorithm that improves the canonicalization of the matrix after the intersection operation which has a time complexity of $O(n^{2})$

\begin{defn} \label{delay}
(The delay operation). Elapsing time means that the upper bounds of the clocks are set to infinity.
That is, after that operation $\forall_{x \in X} : x-x_0 < \infty$ holds. Let $	D' = D\Uparrow$, then:
$$
D^{'}_{i, j} = 
\begin{cases}
 (\infty, <) &  \textrm{for any $i \neq 0$ and $j =0$}, \\
  (D_{i,j})  & \textrm{if $i =0$ or $j \neq 0$} \\
\end{cases}
$$
\end{defn}

The property that all the clocks advance with the same amount of time is ensured by the fact that the constraints on the differences between clocks are not altered by the operation. 
However, the time elapse operation does not break the canonical form of the matrix.

\begin{lem} \cite{Bengtsson04}
The time elapse operation does not break the canonical form of the matrix.
\end{lem}

Recall that when the delay operation is applied it sets all the entries in the first column of the matrix to $(\infty, <)$ and hence these entries will not be changed during canonicalization regardless of the weights of the other constraints in the matrix. This is due to the fact that all acyclic paths of constraints in the matrix that can be used to tighten a constraint $D_{i, 0}$ will pass through a constraint of the form $D_{j, 0}$, where $j \neq i$, and since $D_{j, 0}$ is infinity then $D_{i, 0}$ will remain $\infty$ after canonicalization and hence no need to recanonicalize $D_{i, 0}$. Therefore the matrix that results from opening a zone up will be on canonical form.

\begin{defn} \label{reset}
(The Reset operation). With the reset operation, the values of clocks can be set to zero.
Let $\lambda$ be the set of clocks that should be reset. We can define $D^{'} = D[\lambda:=0]$ as follows.
$$
D^{'}_{j, k} = 
\begin{cases}
(0, \leq) &  \textrm{if $x_j \in \lambda$ and $x_k \in \lambda$}, \\
D_{0, k} &   \textrm{if $x_j \in \lambda$ and $x_k \not \in \lambda$}, \\
D_{j, 0} &   \textrm{if $x_j \not \in \lambda$ and $x_k  \in \lambda$}, \\
D_{j, k} &   \textrm{if $x_j \not \in \lambda$ and $x_k  \not\in \lambda$} \\
\end{cases}
$$

\end{defn}

The reset operation assumes that the given matrix $D$ is in canonical form.
However, the resulting matrix $D^{'}$ may not be in canonical form. 
Note that it is easy to adjust the result to its canonical from, without applying the Floyd-Warshall algorithm, as follows. For each clock $x_i$, reset by this operation, we copy the 0-th column of the DBM to
the $i$-th column, and the  0-th row to the $i$-th row. The resulting matrix
will be in canonical form \cite{Yovine98modelchecking}. 


\section{Behrmann et al. minimum cost reachability algorithm} \label{sec:Behr}

The minimum cost reachability algorithm described in \cite{Behrmann2001} uses a variant of timed automata called  ``uniformly priced timed automata'' and the DBM data structure to 
compute the minimum cost of reaching a goal state in the model.  
A priced timed automata can associate costs with locations, where the costs are multiplied by the amount of time spent in a location.
An automaton may be designed so that the total cost corresponds to the execution time, and thus this approach may be used to calculate the best case execution time problem.
However, the authors in \cite{Behrmann2001} did not give 
a (detailed) formal description of how they extrapolate and canonicalize priced zones of a constructed priced zone graph when they add an extra clock (which they call $\delta$), except the following remark given at page 9.

\begin{quotation}
Termination is ensured if all clocks except for $\delta$ are normalized with respect to a maximum constant $M$. It is important that normalisation never touches $\delta$. With this modification, the algorithm in Fig. 1 will essentially encounter the same states as the traditional forward state-space exploration algorithm for timed automata, 
except for the addition of $\delta$.
\end{quotation}

The remark above does not constitute a concrete definition of extrapolation (normalisation) and does not mention anything about canonicalization, and hence leaves a number of questions concerning implementation open. 
For example, which set of constraints in reachable zones should not be extrapolated? 
How should reachable zones be canonicalized so that correctness and termination of the analysis are guaranteed? 
Do we need to apply different canonicalization procedures on the two sets of constraints in zones: the set of extrapolated constraints and the set of non-extrapolated constraints? More precisely, how the \textit{partially extrapolated zones} (i.e. zones that contain extrapolated and non-extrapolated constraints) should be canonicalized during the analysis? 
Note that key operations of the zone abstraction are canonicalization and extrapolation. Canonicalization assigns the tightest possible bound for each pair of clocks whereas extrapolation enlarges bounds that exceed a certain value after which the value of a clock has no effect on the structure of the zone graph. Canonicalization is needed for the comparison of zones and for efficient implementation of several constraint operations, extrapolation guarantees the finiteness of the zone graph.  
 
However, the extrapolation and canonicalization procedures and their roles in forward reachability algorithm with respect to certain problems such as the minimum termination time problem and the maximum termination time problem, require extra care and non-trivial arguments for proving both correctness and termination in particular when partially extrapolated DBMs.

Suppose we have an automaton $\mathcal{A}$ that we would like to compute 
its minimum and maximum termination times. Suppose further that $\mathcal{A}$ has two clocks
$y$ and $z$. Let us assume that we add an extra clock $\delta$ that is used to compute the execution time of $\mathcal{A}$, which is not reset and not extrapolated during the analysis. 
We can then describe the general form of the partial extrapolated matrix (zone) 
that can be obtained at each reachable location of $\mathcal{A}$ as follows.

\begin{center}
$
M_{PE} = 
\begin{pmatrix}
 & x_{0}  &  y &   z & \delta   \\
x_{0} & . & . & . & * \\
y & . & . & . & * \\
z & . & . & . & * \\
\delta  & * & * & * & * 
\end{pmatrix}
$
\end{center}

where the asterisk sign $(*)$ is used to denote a constraint involving $\delta$ which is a constraint that is not changed during extrapolation and the dot sign $(.)$ is used to denote a constraint involving only the automaton clocks which may be changed during extrapolation and it is in $M$-form. We say that a constraint $(D_{i, j}, \prec_{i,j})$ is in $M$-form if $-M \leq D_{i, j} \leq M$.
As one can see all the constraints involving the extra clock $\delta$ (the asterisk entries) do not change during extrapolation in order to keep the extra clock precise during the construction of the zone graph. On the other hand, the constraints involving the automaton clocks (the dot entries) may be changed during extrapolation in order to guarantee termination in particular when infinite cycles exist.
Note that during the construction of the zone graph $Z(\mathcal{A})$ the only operation in which the set of non-extrapolated constraints may influence the set of extrapolated constraints and vice versa 
is the canonicalization operation where the constraints in the matrix are tightened (see Section  \ref{sec:ZoneAppraoch}). Note that canonicalization may be repeated several times at each step of the successor computation.  Hence, canonicalization needs to be performed carefully so that
the constraints involving the extra clock $\delta$ remain precise while guaranteeing termination of the analysis.

\begin{figure}
    \centering
    \includegraphics[width= 3in]{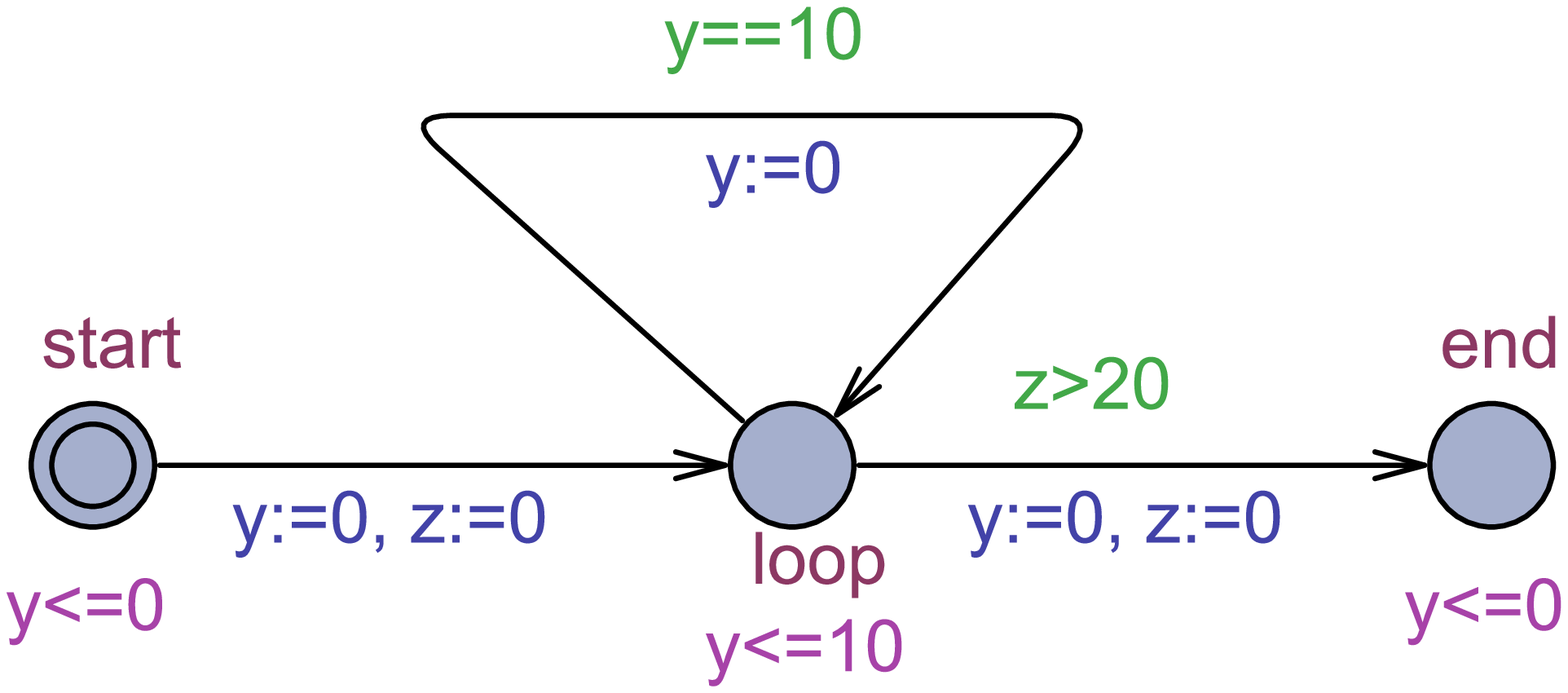}
    \caption{An automaton with BCET = 21 and WCET =$\infty$}
    \label{fig:infinite}
    \centering
    \includegraphics[width= 3in]{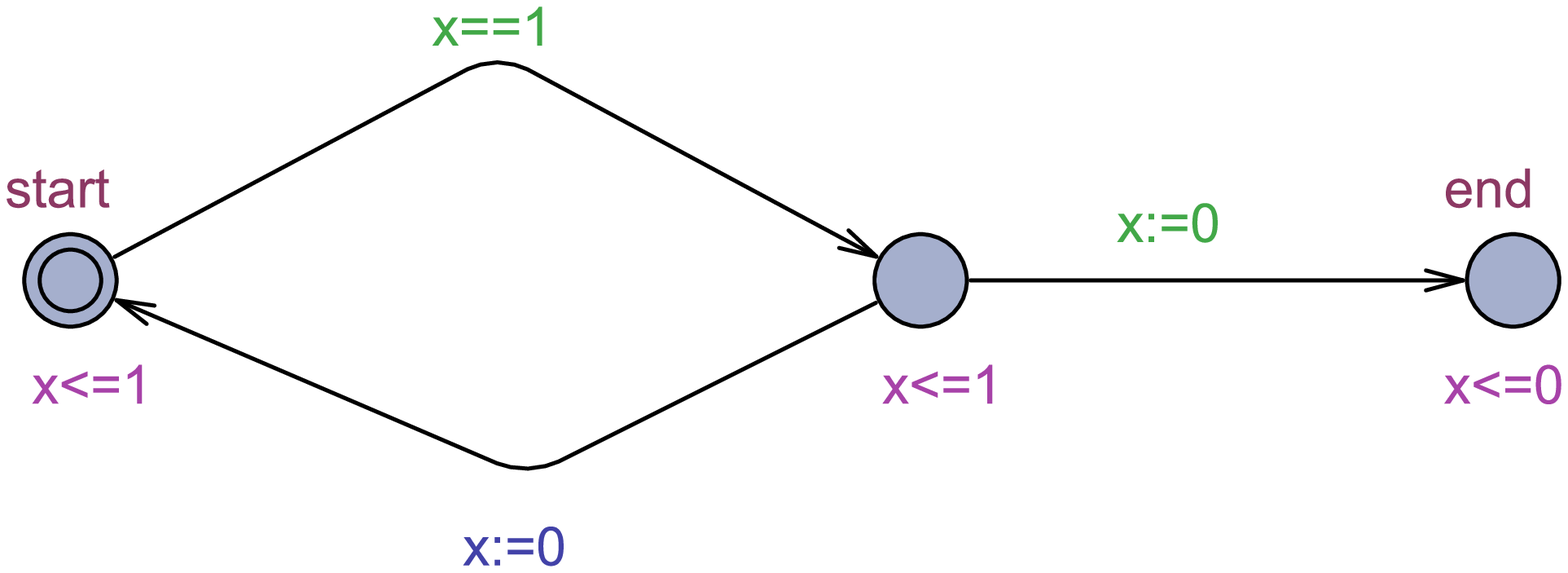}
    \caption{An automaton with BCET =1 and WCET =$\infty$}
    \label{fig:simpleInfinite}
\end{figure}

\begin{figure} [h!]
    \centering
    \includegraphics[width= 3in]{infiniteone}
    \caption{ An automaton with a BCET =11 and WCET =$\infty$}
    \label{fig:infiniteSecond}
    \centering
    \includegraphics[width= 3in]{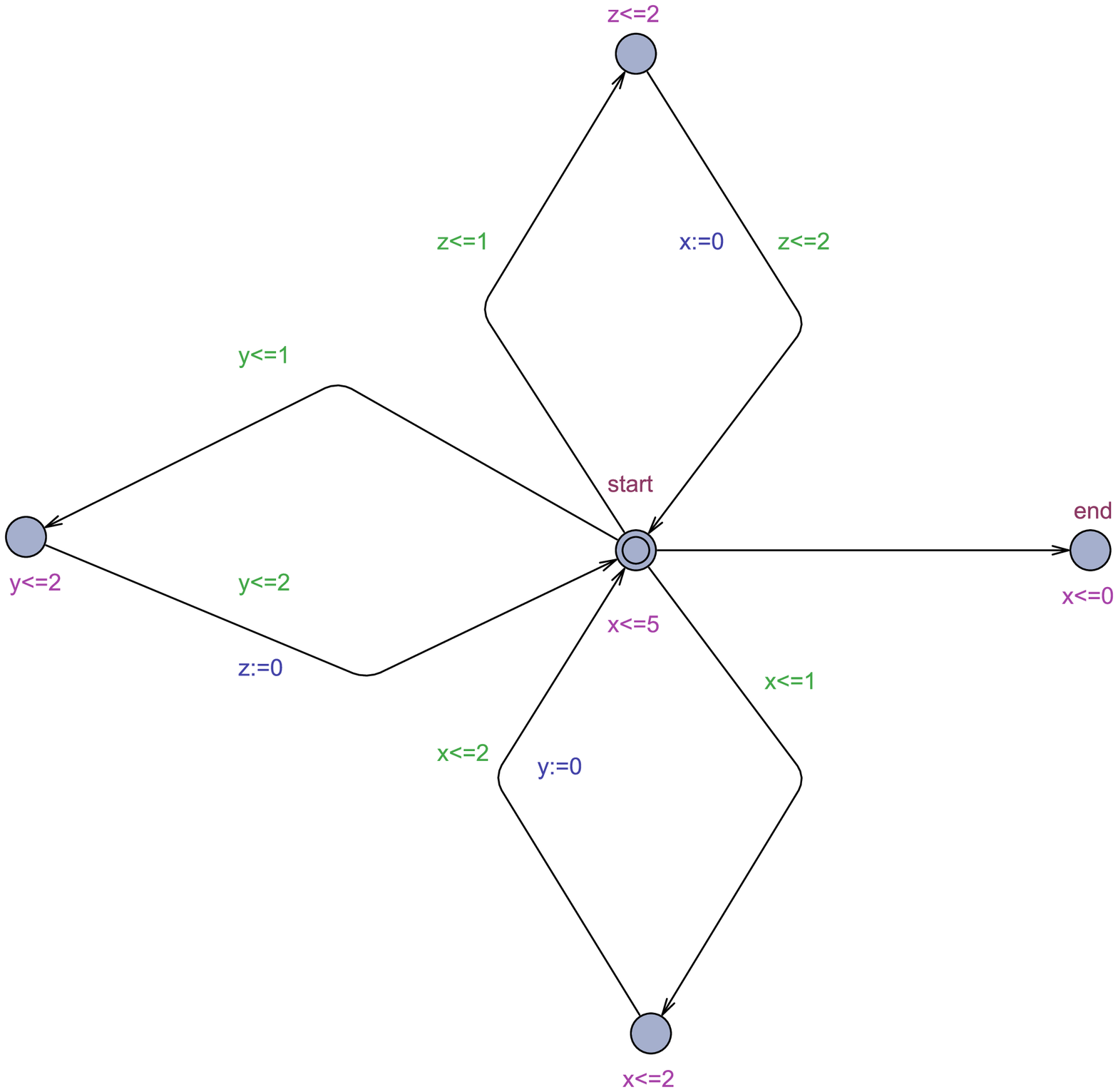}
    \caption{ An automaton where  BCET =0 and WCET =$\infty$ }
    \label{fig:infiniteWithsThreeFinite}
\end{figure}

Before discussing the special conditions needed to define a forward zone-based reachability
algorithm that terminates and computes the correct maximal time,  we show first that in certain circumstances, when infinite cycles exist, the algorithm in \cite{Behrmann2001} and its implementation in UPPAAL may not terminate. To support our claim we give four examples of TA where the algorithm as described in \cite{Behrmann2001} does not guarantee termination. However, to support further our theoretical claim we give the results of verifying the examples using the latest version of the tool UPPAAL (4.1.19), Windows version, which show that the tool fails to terminate. 
Note that in UPPAAL, one can use a global clock GBL and check two properties on system A:
(\verb+inf+ \{ A.end \} : GBL)  and  (\verb+sup+ \{ A.end \} : GBL).
The sup/inf operators are documented in the Help menu of UPPAAL.
The example in Figure \ref{fig:infinite} shows an automaton with an infinite cycle (loop) where  BCET is 21 and WCET is infinity. UPPAAL fails to give an answer for BCET and WCET of that automaton. UPPAAL also fails to handle the simple infinite automaton given in Figure \ref{fig:simpleInfinite} where BCET is 1 and WCET is infinity. For this particular automaton we verify the BCET using the command (\verb+inf+ \{ A.end \} : GBL) and the WCET using the command (\verb+sup+: GBL). However, UPPAAL fails to terminate and hence no answer has been obtained. For the automaton in Figure \ref{fig:infiniteSecond} which has a BCET of 11 and an infinite WCET UPPAAL fails also to handle this automaton and no answer has been obtained for both BCET and WCET.
On the other hand, the example in Figure \ref{fig:infiniteWithsThreeFinite} shows an automaton that contains three finite cycles that have the location \verb+start+ as a common location. It is interesting to note that there are some dependencies between the behaviour of the three cycles. However, as one can see, the three cycles collectively will be executed infinitely often which lead to an infinite WCET. UPPAAL fails to handle such classes of cycles where the operator \verb+sup+ fails to terminate and hence no answer has been obtained.

Let us see what happens when we compute the zones of the automaton in Figure \ref{fig:infinite} using the partial $M$-extrapolation algorithm proposed by Behrmann et al in \cite{Behrmann2001} where the extra clock is not touched during extrapolation. 
Firstly, note that the extrapolation constant $M$ is 20. The automaton has two clocks $y$ and $z$.
Let us call the extra clock $\delta$. We give the sequence of zones obtained below. 
Note that for convenience only the full canonical zone is written.  
First at location $\verb+start+$ we have the zone $(\delta =0 \land y =0 \land z =0)$. 
During the forward traversal of the TA the location \verb+loop+ is reached with the clock zone 
$ (\delta \leq 10 \land y =0 \land \delta  = z)$. Clearly, extrapolation is not necessary here
since none of the constraints exceeds the extrapolation constant $M$.
After taking the transition $\verb+loop+ \rightarrow \verb+loop+ $ a state $(\verb+loop+ , Z_{2})$
with $Z_2 = (\delta \leq 20 \land y =0 \land \delta  = z)$ will be added. Again extrapolation is not necessary here.
A second loop will add  $(\verb+loop+ , Z_{3})$ with $Z_3 = (\delta \leq 30 \land y =0 \land \delta  = z)$.
Before proceeding further, note  that the zone $Z_3$ needs to be extrapolated since
there are some constraints that exceed the value of the extrapolation constant $M$. 
Recall that in the partial extrapolation approach we do not change all the asterisk entries in the matrix (i.e. the entries involving $\delta$) in order to keep them precise and we just extrapolate the dot entries (i.e. the entries involving the automaton clocks). One can check this would give the zone $Z_{3}^{'} = (\delta \leq 30 \land y =0 \land \delta - z =0 \land z = \infty)$. Note that the zone $Z_{3}^{'}$ is not on a canonical form.                                                                                                                       We use the Floyd-Warshall algorithm to canonicalize the zone $Z_{3}^{'}$. 
We obtain the zone $Z_{3}^{''} = (\delta \leq 30 \land y =0 \land z \leq 30)$ 
which is the one we obtained before extrapolation. 
If we continue computing the zones in this way we will find ourselves dealing with real zones rather than abstracted ones and hence the loop can be taken infinitely often enlarging the state space such that a fixed point will never be reached. This happens because the constraints involving $\delta$ have not been changed during extrapolation and then during the canonicalization step the value of these constraints influenced 
the value of the constraints involving the automaton clocks. 
This explains why verification of the above four examples in UPPAAL does not terminate and that no answer can be obtained in such cases!

\section{Computing WCET of Cyclic Real-time Systems}

An algorithmic solution to the WCET problem proceeds by adding an extra	
clock (let us call it $\delta$) to the automaton under analysis 
that acts as an observer. Then one computes the zone graph of the automaton (involving $\delta$), by means of a standard forward analysis using DBMs. To get the WCET, the algorithm needs to look at the value of the constraint $ (D_{i, 0}, \prec_{i, 0})$ in every reachable state including the initial state since the delay of a run can be infinity if there is an unconstrained location along that run or if there is an infinite cycle in which time can elapse.
 
\subsection{Solving The Problem Using Partial Extrapolation} \label{sec:partialEXtra1}
 
We discuss now an extrapolation procedure that can be used to keep the extra clock precise to the end of the analysis. We use the term ``partial extrapolation'' for such a procedure.
Let us denote the sub-DBM that consists in the asterisk entries as  $\overset{*}{M_{PE}}$, 
and the sub-DBM that consists in the dot entries as $\dot{M_{PE}}$ which may be extrapolated during the extrapolation steps. So to solve the problem we choose to split the DBM $M_{PE}$ into two sub-DBMs 
$\overset{*}{M_{PE}}$ and $\dot{M_{PE}}$. Note that such splitting is possible since DBMs are sets of constraints. We give now the conditions that are necessary to ensure correctness and termination of the analysis using the partial extrapolation approach. 

\begin{enumerate}

\item  (\textbf{Condition C1: special extrapolation procedure of reachable zones}). 
During the extrapolation steps, extrapolate only
the dot entries in the matrix and leave all the asterisk entries unextrapolated.  Note that this is necessary in order to keep the constraints involving the extra clock precise to the end of the analysis. For greater convenience we will use the notation $\verb+MExtra+_{M}(D)$ to denote the  modified $M$-extrapolation operation and to distinguish it from the classical $M$-extrapolation operation (see Definition 6). Note that in the operation $\verb+MExtra+_{M}(D)$ all the asterisk entries are not changed. Let us assume that the clock $\delta$ takes index $i$ in DBMs. We can then compute the operation $\verb+MExtra+_{M}(D)$ of a given matrix $D = (d_{j,k}, \prec_{j,k})_{j,k=0,..n}$ as follows.
$$
(d^{'}_{j,k}, \prec^{'}_{j,k}) = 
\begin{cases}
 (\infty, <) &  \textrm{if $d_{j,k}> M \land j, k \neq i$}, \\
  (-M, <) & \textrm{ if $d_{i,j} <-M \land j, k \neq i$}, \\
  (d_{j,k}, \prec_{j,k})  & \textrm{otherwise.} \\
\end{cases}
$$

\item  (\textbf{Condition C2: special canonicalization procedure for dot entries}). 
The dot entries in the matrix need to be canonicalized independently or separately from the asterisk entries. That is, during the canonicalization steps, the asterisk entries
should not participate in the process of canonicalizing the dot entries.  Otherwise, termination may not be guaranteed. To see why condition C2 is necessary, consider the case where extrapolating a constraint $(d_{j, k}, \prec_{j,k}) \in \dot{M_{PE}}$ yields $(\infty, <)$. 
Now if the asterisk entries participate in the process of canonicalizing the constraint $(\infty, <)$ it is possible to end up with a  constraint $(d^{'}_{j, k}, \prec^{'}_{j,k})$ which may not be in $M$-form (i.e. extrapolated form) in the sense that $d^{'}_{j, k} > M$ or $d^{'}_{j, k} < - M$. We already observed this when discussing Behrmann et al minimum cost reachability algorithm in the previous section (see the analysis of the example at Figure \ref{fig:infinite} which leads to non-termination when using Behrmann et al partial $M$-extrapolation technique).


\item  (\textbf{Condition C3:  canonicalizing asterisk entries in the matrix}). 
For the asterisk entries in the matrix it is necessary to canonicalize them
using dot and asterisk entries. That is, to canonicalize the asterisk entries using the classical canonicalization procedure. 
This ensures that the extra clock will advance at the same rate as the automaton clocks.
Note that although the constraints involving the extra clock will
be canonicalized using the entire set of constraints in the zones including the extrapolated constraints, 
the extra clock does not lose its precision in the end as one might expect.
The reason is that during the construction of a zone graph the zones are canonicalized before they get extrapolated and that during canonicalization, 
a minimum is calculated. When extrapolation is applied it only increases
bounds in the zone and that the extra clock is not changed during extrapolation. Moreover, the relationship of the extra clock to the other automaton clocks are preserved by partial extrapolation  where all the constraints involving the extra clock will not be changed during extrapolation (see condition C1). The key idea is that increasing bounds during extrapolation does not affect a function that was calculated using a minimum.  For greater convenience we will use the notation \verb+MCanon(D)+ to denote the modified canonicalization procedure and to distinguish it from the standard canonicalization operation. Note that the operation \verb+MCanon(D)+ is a specialisation of Floyd's algorithm in which the dot entries are tightened separately from the asterisk entries,  while the asterisk entries are tightened using asterisk and dot entries as shown in Algorithm \ref{alg:Floyd-WarshallChp4}.

\item (\textbf{Condition C4: special checks for handling cycles in TA}). 
To handle cycles (loops) properly we propose to use what we call fixed point abstraction of zones (see Definition \ref{FP1}) rather than inclusion abstraction when handling the generated zones inside cycles. That is, during the analysis of a cycle in a TA we check whether the search can reach identical states with respect to the automaton clocks and whether the extra clock (i.e. non-extrapolated clock) can advance during the analysis of the cycle. If such situation happens we set the upper bound  of the extra clock $\delta$ to infinity and terminate since the WCET of the automaton will be infinity. Note that reaching a fixed point of infinite cycles in TA is guaranteed as zones are extrapolated during the analysis (see Lemma \ref{fixedpointTheorem}).

\end{enumerate}

\begin{defn} (\textbf{Fixed Point of a cycle})\label{FP1}.
Let $\mathcal{A} =  (\Sigma, L, L_{0}, L_{F}, X, I, E)$ be a timed automaton, let $E_{\pi} = (e_0, ...,e_{n-1})$ be the sequence of edges of a cycle $\pi$ in $\mathcal{A}$. Suppose that the operation $succ(Z, E_{\pi})$ computes the successor zone of $Z$ after executing the sequence of edges in $E_{\pi}$ which is equivalent to executing the cycle $\pi$ one full iteration. We say that $Z$ is a fixed point of $\pi$ and $\pi$ is an infinite cycle if  $succ(Z, E_{\pi}) = Z$. That is, if the cycle starts and ends with the same zone then the cycle is an infinite cycle.

\end{defn}

\begin{lem} \label{fixedpointTheorem}
Let $\pi$ be a cycle in an automaton $\mathcal{A}$ that can be run infinitely often.  Then after a finite number of iterations a fixed point of $\pi$ will be reached given that the $M$-extrapolation operation is applied during the analysis.
\end{lem}


Conditions C2 and C3 in the above described procedure may not be straightforward conditions as the other ones so it may be worth providing some formal argument why these conditions are necessary for the correctness of the procedure. To explain formally why conditions C2 and C3 are necessary we need some preliminary observations. The first observation is that extrapolation only increases bounds, never decreases them. The second observation is that canonicalization only decreases bounds, never increases them. The third observation is that the constraints involving the extra clock $\delta$ are not touched (enlarged) during extrapolation. From these three observations it is easy to see that if a constraint $c$ in a canonical zone $Z$ has not been touched during extrapolation then the weight of $c$ in the canonicalized extrapolated matrix can not be smaller than its weight in the canonicalized non-extrapolated matrix since extrapolation increases bounds and never lowers them (i.e. $Z \subseteq \verb+MExtra+(Z)$). 
The above observations lead to lemma  \ref{optLemma1} which is interesting since it discusses a result that has not been noticed in the prior literature.

\begin{lem} \label{optLemma1}
Let $Z$ be a zone on canonical form. Let $(c_{i, j}, \prec_{i, j})$ be a constraint in $Z$.
Suppose that the zone $Z$ has been partially extrapolated using the $M$-extrapolation procedure (see Definition \ref{m-norm}) and that the constraint $(c_{i, j}, \prec_{i, j})$ has not been touched during extrapolation. Then the weight of $(c_{i, j}, \prec_{i, j})$ in the canonical matrix $Z$ is equal to its weight in the matrix $\verb+MCanon+ (\verb+MExtra+_{M}(Z))$ and hence $(c_{i, j}, \prec_{i, j})$ needs not to be recanonicalized after extrapolation.
\end{lem}

Lemma \ref{optLemma1} explains to us why condition C3 in the above procedure is sound and why the canonicalization operation at step 8 of the zone approach (see Section \ref{sec:ZoneAppraoch}) will not affect adversely the constraints involving the extra clock $\delta$ (the asterisk entries) and hence they remain precise. In fact, the asterisk entries will not be changed at step 8 of the zone approach. From the above observations it is easy to see also that condition C2 is necessary since the dot entries may be increased during extrapolation and then during the canonicalization operation the asterisk entries may influence the dot entries in a way they may lose their $M$-form. This can affect adversely termination of the analysis in particular when infinite cycles exist (see examples in Section \ref{sec:Behr}).

Conditions C2 and C3 can be formalized as described in Algorithm \ref{alg:Floyd-WarshallChp4} where the asterisk entries in the matrix are canonicalized using asterisk and dot entries while the dot entries are canonicalized using only dot entries. 
Recall that we assume that the clock $\delta$ takes index $i$ in DBMs.

\begin{algorithm} 
\caption{Special canonicalization procedure at steps (4, 7) of the zone approach}
\textbf{for} $p := 0$ \textbf{to} $n$ \textbf{do} \\
\hspace*{10pt} \textbf{for} $q := 0$ \textbf{to} $n$ \textbf{do} \\
\hspace*{20pt} \textbf{if} $p = i \lor q = i$ \textbf{then} \\ 
\hspace*{30pt} \textbf{for} $k =0$  \textbf{to} $n$ \textbf{do} \\
\hspace*{40pt}  $D_{p, q} := \min (D_{p, q},  D_{p, k} + D_{k, q})$ \\
\hspace*{30pt} \textbf{end} \\
\hspace*{20pt} \textbf{else} \\
\hspace*{30pt} \textbf{for} $k =0$  \textbf{to} $n$ \textbf{do} \\
\hspace*{40pt} \textbf{if} $k \neq i$ \textbf{then}  $D_{p, q} := \min (D_{p, q},  D_{p, k} + D_{k, q})$ \\
\hspace*{30pt} \textbf{end} \\ 
\hspace*{10pt} \textbf{end} \\
\textbf{end}
  \label{alg:Floyd-WarshallChp4}
\end{algorithm}

An advantage of Lemma \ref{optLemma1} is that it leads to the optimised canonicalization procedure described in Algorithm \ref{alg:SpecOfFloyd-Warshall1}, which states that after extrapolation only the constraints that have been changed  need to be recanonicalized. Note that the list \verb+Changed+ maintains the list of constraints that have been changed during extrapolation represented as pairs of indices. Recall that all the dot entries in the matrix need to be canonicalized separately from the asterisk entries (see Condition C2).

\begin{algorithm}
\caption{Special canonicalization procedure after extrapolation} 
\label{alg:SpecOfFloyd-Warshall1}
\textbf{for} $k := 0$ \textbf{to} $n$ \textbf{do} \\
\hspace*{10 pt} \textbf{if} $k \neq i$  \textbf{then} \\ 
\hspace*{20 pt} \textbf{for} $(p, q) \in \verb+Changed+$  \textbf{do} \\
 \hspace*{30 pt} $D_{p,q} := \min (D_{p,q}, D_{p,k} + D_{k,q} )$ \\
\hspace*{20 pt} \textbf{end} \\
\textbf{end}
\end{algorithm}

One may argue that time elapse operation (see Definition \ref{delay}) can affect the upper bound of $\delta$ in a way it becomes imprecise. Note that time elapse does only affect the upper bound of $\delta$, but not it's relationship with other clocks. After time elapse, if there is just one other clock whose upper bound is not infinity
due to extrapolation, the relationship of $\delta$ to this clock is preserved and thus the exact upper
bound of $\delta$ can be reconstructed during canonicalization. During canonicalization, a minimum is calculated,  thus the smallest upper bound dominates all others. 
However, if all the upper bounds of the clocks are set to infinity by extrapolation  
so that none of the clocks remains tight after extrapolation, then there is a path in the automaton
that has $\mathit{WCET} = \infty$, as all clocks used in guards or invariants are beyond the biggest constant they ever compared against, thus the automaton must be in a state without an upper bound of the location.

It remains to discuss how the procedure works in the presence of finite cycles (i.e. cycles that can be repeated a finite number of times). Note that for finite cycle the search will not reach a fixed point 
but there will be an iteration of the cycle where the search encounters a blocking clock and hence the cycle can not be repeated any further.  However, since we seek a solution to the problem in the presence of extrapolation
it is necessary then to ensure that the search does not leave finite cycles before executing them the precise number of times. More concretely, we need to ensure that if a finite cycle can be repeated $n$ times in the non-extrapolated graph, where $n < \infty$, then it can be repeated also $n$ times in the extrapolated graph
and that the minimum and maximum total execution time of the finite cycle in the extrapolated graph are equal to those obtained in the non-extrapolated graph. This is what we show in Corollary \ref{finiteCycleTheorem}.

\begin{thm} \cite{Bengtsson04} Let $(l_0, Z_0)$ be an initial state of an automaton $\mathcal{A}$
where $l_0\in L_0$ and $Z_0$ is the corresponding initial clock zone.
Let $M = \max(\mathcal{A})$ be the maximal integer that appears in the guards and the location invariants of $\mathcal{A}$ and $\rimp_{M}$ be the transitions resulting from the $M$-extrapolation.
Let $\mathcal{B}(X)$ be the set of logical formulae generated by the syntax $g:= y \sim c \mid g \land g$ where $\sim \in \{<, \leq \}$.
Assume that $D^f \in \mathcal{B}(X)$.
\begin{itemize}

\item (Soundness) whenever $(l_0, Z_0) \rimp_{M} (l_f, Z_f)$ then   $(l_0, Z_0) \rimp (l_f, Z_f)$ for all
$Z_{f} \in D^f$

\item (Completeness) whenever $(l_0, Z_0) \rimp (l_f, Z_f)$ then   $(l_0, Z_0) \rimp_{M} (l_f, Z_f)$ for all
$Z_{f} \in D^f$

\end{itemize}

\end{thm}


\begin{cor} \label{finiteCycleTheorem}
Let  $\pi$  be a cycle in an automaton $\mathcal{A}$. 
If $\pi$ can be repeated $n$ times in the graph $Z(\mathcal{A})$
then $\pi$ can be repeated also $n$ times in the graph $\verb+MExtra+_M(Z(\mathcal{A}))$ and that the lower and upper total delays of $\pi$ in $Z(\mathcal{A})$ are equal to those obtained in $\verb+MExtra+_M(Z(\mathcal{A}))$.
\end{cor}

The modified zone based approach with conditions C1-C4 may yield zones that are partially extrapolated and partially canonicalized. Since the asterisk entries in the matrix will not be touched during extrapolation and that during canonicalization the dot entries will be canonicalized using only dot entries while the asterisk entries will be canonicalized using dot and asterisk entries. This is necessary for the correctness and the termination of the WCET analysis since the observable clock $\delta$ that does not interfere with the guards of the automaton must not be touched during extrapolation and has to advance at the same rate as the automaton clocks. 
However, the procedure guarantees termination since there is a finite number of sub-DBM $\dot{M_{PE}}$ due to conditions C1, C2, and C4. The procedure also keeps information precise for $\delta$ due to conditions C1, C3, and C4. This is what we prove in Theorem \ref{partial-extrapolation}.

\begin{thm} \label{partial-extrapolation}
The partial extrapolation algorithm that satisfies conditions (C1-C4) keeps the observable clock $\delta$ precise to the end (i.e. the clock $\delta$ preserves its actual value) and the algorithm guarantees termination. 
\end{thm}

\section{A Zone-based Algorithm for Computing WCET of TA} \label{sec: algorithm}

Algorithm \ref{alg:WCETAlgorithm}
gives a zone-based algorithm for calculating the WCET of real-time distributed systems. 
The algorithm takes as input an automaton $\mathcal{A}$ for the system to be analysed.
Each node in the computed tree is of the form $(l_{i}, \verb+MExtra+_{M}(Z_{i}), sts)$ where $l_{i}$ is a location in the automaton,
$Z_{i}$ is the corresponding partially extrapolated zone, and $sts$ is an integer variable which is assigned to each state
in order to detect whether there exists a cycle on locations in the behaviour of the automaton.
The variable $sts$ can take values from the set $\{0, 1, 2 \}$. When it is 0 it means that the location 
has not been visited before, when it is 1 it means the location has been visited before but not fully explored,
and when it is 2 it means that everything reachable from that location have been explored.
We assume that the reader is familiar with the classical DFS algorithm with the labelling process
of nodes to unvisited (0), being explored (1), and finished (2) and hence we omit these details.
The algorithm uses two data structures WAIT and PASSED to store symbolic states waiting to be examined,
and the states that already examined, respectively.
The WAIT set is instantiated with the initial symbolic state $(l_{0}, Z_{0}, 0)$. 
The global variable \verb+WCET+ holds the currently longest known execution time for reaching the final location; initially it is 0.
The global clock $\delta$ keeps track of the execution time of the system.
In each iteration of the \textbf{while} loop, the algorithm selects a symbolic state $s$ from WAIT, checking if the state is a final state. 
If the state does not evolve to any new state then we consider it as a final state of some branch in the graph.
If the state $s$ is a final state we update the best known \verb+WCET+
to the upper bound value of $\delta$ at $s$ if it is greater than the current value of \verb+WCET+. 
If the state is not a final state, we add all successors of $s$ to WAIT and continue to the next iteration.
During the search, if the algorithm encounters  a reachable location that is not guarded by an invariant 
the search can stop immediately since the WCET will be infinity. 
Similarly, if the search detects an infinite cycle in the automaton at which time can elapse then it stops immediately since the WCET will be infinity.
It is interesting to note that the extrapolation procedure used in the algorithm is the partial
extrapolation that satisfies conditions C1-C4. Note that we write $s.\dot{Z}$ to refer to the extrapolated sub-DBM $\dot{Z}$ that consists in the dot entries in the matrix $Z$ (i.e. the entries involving the automaton clocks) in the state $s$.

\begin{algorithm} [h]
\caption{An algorithm for computing WCET of diagonal-free TA}
\label{alg:WCETAlgorithm}
\textbf{Input}: $(\mathcal{A}$)\\
\textbf{Output}:  $\verb+WCET+ := 0$\\
\textbf{clock} $\delta$ \\
PASSED := $\emptyset$;  WAIT := $\{(l_{0}, Z_{0}, 0) \}$ \\
\textbf{while} WAIT $\neq ~ \emptyset$ \\
\hspace*{5pt} select $s$ from WAIT \\
//Check if $s$ is a final node on some branch of the tree \\
\hspace*{5pt} \textbf{if}  for all $a \in \Sigma$  $post_{a}(s) = \emptyset$ ~ \textbf{then}  \textbf{if} $upperBound (s.Z,  \delta) > \verb+WCET+$ \\ \hspace*{10pt} \textbf{then}  $\verb+WCET+ :=  upperBound (s.Z,  \delta)$ \\
\hspace*{5pt} add $s$ to PASSED \\
\hspace*{5pt} for all $s^{'}$ such that  $s \leadsto s^{'}$ do \\
// if there exists a location that	is not guarded with an invariant \\
\hspace*{10pt}  \textbf{if} $upperBound (s^{'}. Z,  \delta) = \infty $  \textbf{then} $\{ \verb+WCET+ :=\infty;  \textbf{return}  ~ \verb+WCET+ \}$ \\ 
// if there is an infinite cycle in the automaton at which time can elapse \\		
\hspace*{10pt} \textbf{else if} $ s^{'}.l = s^{''}.l \land s^{'}.\dot{Z} = s^{''}. \dot{Z} ~ \land$  $(s^{'}.Z_{\delta, 0} > s^{''}.Z_{\delta, 0}) \land s^{''}.sts = 1$ \\
\hspace*{10pt}  for any $s^{''} \in$ PASSED  \textbf{then}  $\{ \verb+WCET+ :=\infty; \textbf{return}  ~ \verb+WCET+ \}$  \\
// if there is an infinite cycle in the automaton at which time cannot elapse  \\ 
\hspace*{10pt}  \textbf{else if} $ s^{'}.l = s^{''}.l \land s^{'}.Z = s^{''}.Z \land s^{''}.sts = 1 $   for any $s^{''}  \in$ PASSED   \verb+continue+ \\
// if the state $s^{'}$ is a new state \\
\hspace*{10pt} \textbf{else}  add $s^{'}$ to WAIT  \\ 
\textbf{return} \verb+WCET+
\end{algorithm}

\begin{thm} \label{mainThrm}
The zone-based Algorithm \ref{alg:WCETAlgorithm} computes correctly the WCET of any diagonal-free TA  $\mathcal{A}$ and guarantees termination.

\end{thm}

\section{Complexity}

In Table \ref{table:complexity} we summarise the necessary DBM operations used by the algorithms with their complexity. We refer the reader to \cite{Bengtsson04} for more details about how one can compute complexity of each of these operations.  All required operations can be implemented on DBMs with satisfactory efficiency.
Given the time complexity of each DBM operation performed by the algorithms
we end up with a  time complexity of the form given in Theorem \ref{time-complexity}, 
where $d$ is the number of states in the WAIT list that have the same discrete part with the new generated state 
that results from executing the operation $post_{a}(s)$,
we use this for the fixed point test operation.
Note that the value of $d$ is bounded by the number of generated zones ($|Z|$) of the automaton under analysis.


\begin{table}[h]
  \centering
\begin{tabular}{|c|c|c|c|}
\hline
DBM-operation & Complexity  & DBM-operation & Complexity\\
\hline 
Fixed-point test & $O(|X|^{2})$ & Consistency test & $O(|X|^{2})$ \\ 
\hline
Extrapolation & $O(|X|^{2})$ &  Canonicalization &  $O(|X|^{3})$ \\
\hline
Resetting Clocks & $O(|X|)$  & Delay & $O(|X|)$\\
\hline
Constraint intersection & $O(|X|^{2})$ & Clock-upper bound check & $O(1)$ \\
\hline
\end{tabular} 
\caption{Complexity of WCET algorithm in terms of DBM operations \label{table:complexity}}
\end{table}

\begin{thm} \label{time-complexity}
 
The WCET zone-based algorithm has a time complexity of the form $O( (|X|^{3}+ d. |X|^{2}). |E|. |Z|)$, where $|Z|$ is the number of generated zones of the automaton under analysis, $|X|$ is the number of clocks in the automaton, and $|E|$ is the number of reachable edges in the automaton.
 \end{thm}

\section{Implementation}

In this section we briefly summarise our prototype implementation of the model checking
algorithms given in Section \ref{sec: algorithm}. 
It is important to note that the goal of
our implementation is to validate the presented algorithms, rather than to
devise an efficient implementation; this will be the subject of our future work.

The prototype implementation has been developed using the opaal tool \cite{opaal} 
which has been  designed to rapidly prototype new model checking algorithms.
The opaal tool is implemented in Python and is a standalone model checking engine.
Models are specified using the UPPAAL XML format.
We use the open source UPPAAL DBM library for the internal symbolic representation of time zones in the algorithms.

We consider here a simple realistic automatic manufacturing
plant taken from Daws and Yovine \cite{Daws1995}. 
We first give an informal description of the case study 
then we give the timed automata model of the entire system in UPPAAL,
and finally report on the results obtained from running the BCET/WCET algorithms
on the case study when considering it under different configurations.

The manufacturing plant that we consider consists of a conveyor belt that moves from left to
right, a processing or service station, and two robots that move boxes between
the station and the belt. The first robot called D-Robot takes a box from the station and put it
on the left end of the belt. The second robot called G-Robot picks the box from the right end of the
belt and transfers it to the station to be processed. We are then interested in verifying the minimum
and maximum amount of time a box can take to be processed when considering the manufacturing plant under different configurations.

\begin{figure} 
  \begin{minipage}[b]{0.5\linewidth}
    \centering
    \includegraphics[width= 2.3in]{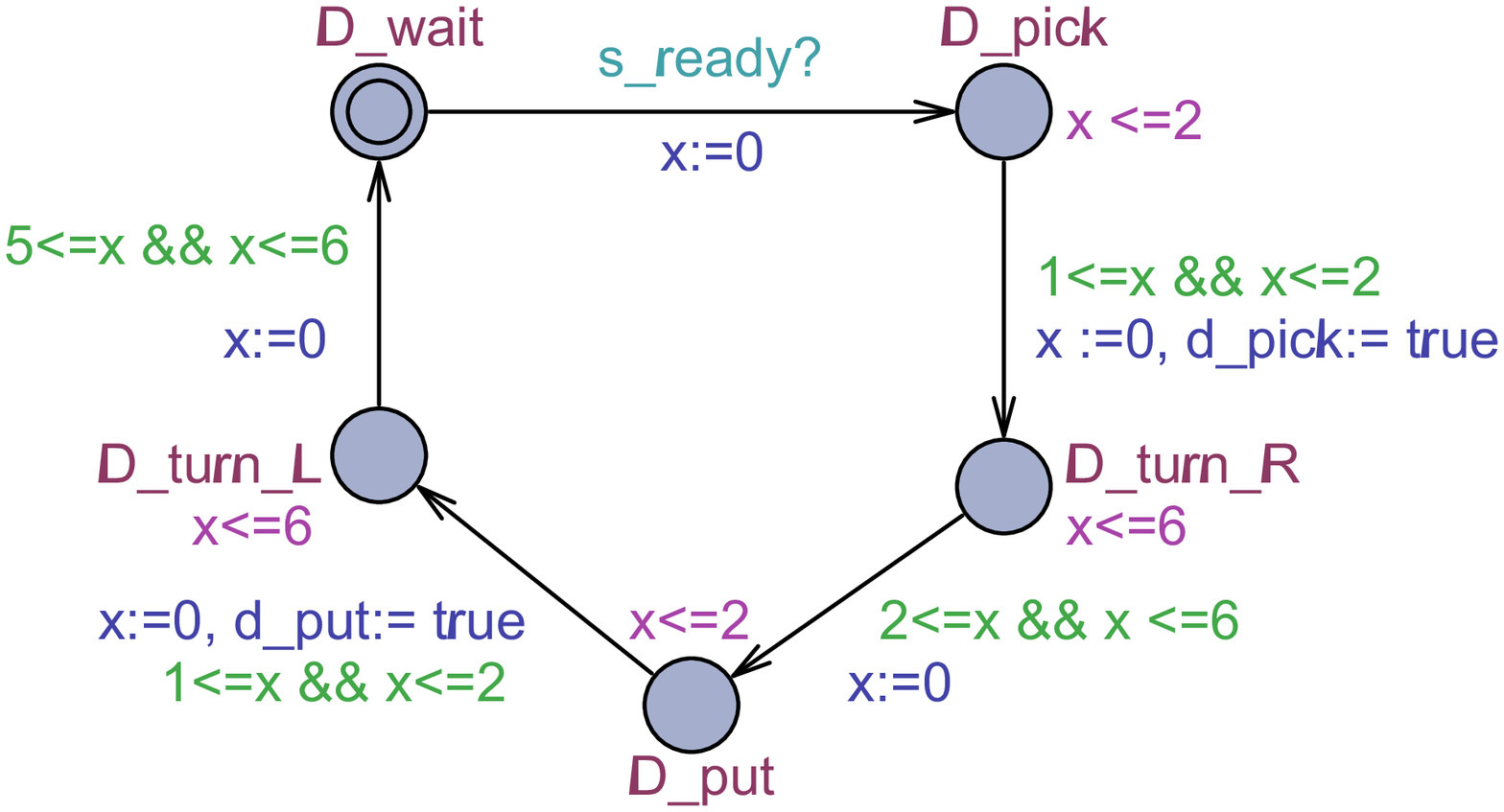}
    \caption{The D-Robot template}
    \label{fig:D-Robot}
  \end{minipage}
  \hspace{0.5cm}
  \begin{minipage}[b]{0.4\linewidth}
    \centering
    \includegraphics[width= 2.3in]{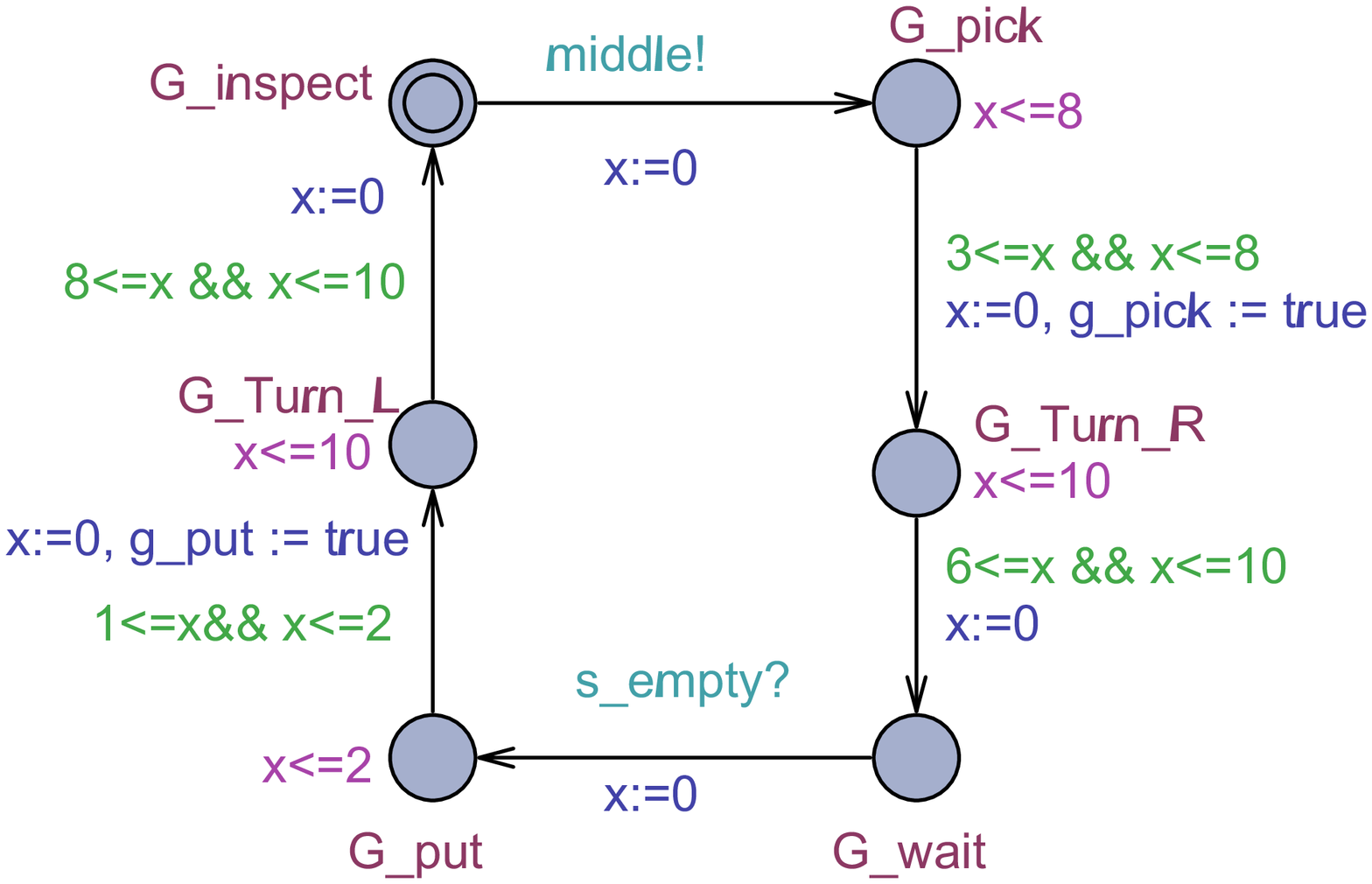}
    \caption{The G-Robot template}
    \label{fig:G-Robot}
  \end{minipage}
\end{figure}

The timed automaton for the D-Robot is given in Figure \ref{fig:D-Robot}. Initially,
the robot waits until a box is ready indicated by the synchronisation label \verb+s-ready+.
Next, it picks the box up, turns right and puts the box on the moving belt.
It then turns left and returns to its initial position.

The timed automaton for the G-Robot is given in Figure \ref{fig:G-Robot}.
This robot waits at the inspection point at the right end of the belt until
a box passes this point. The G-Robot must pick up the box before it falls off
the end of the belt. Next, it turns right, waits for the station to finish processing
the previous box and then puts the box at the station. 
Finally, it turns left back to the inspection point.
Note that picking the box up by the robot,
turning left or right takes time which depends mainly
on the speed of the robot.

\begin{figure}
  \begin{minipage}[b]{0.5\linewidth}
    \centering
    \includegraphics[width= 2.4in]{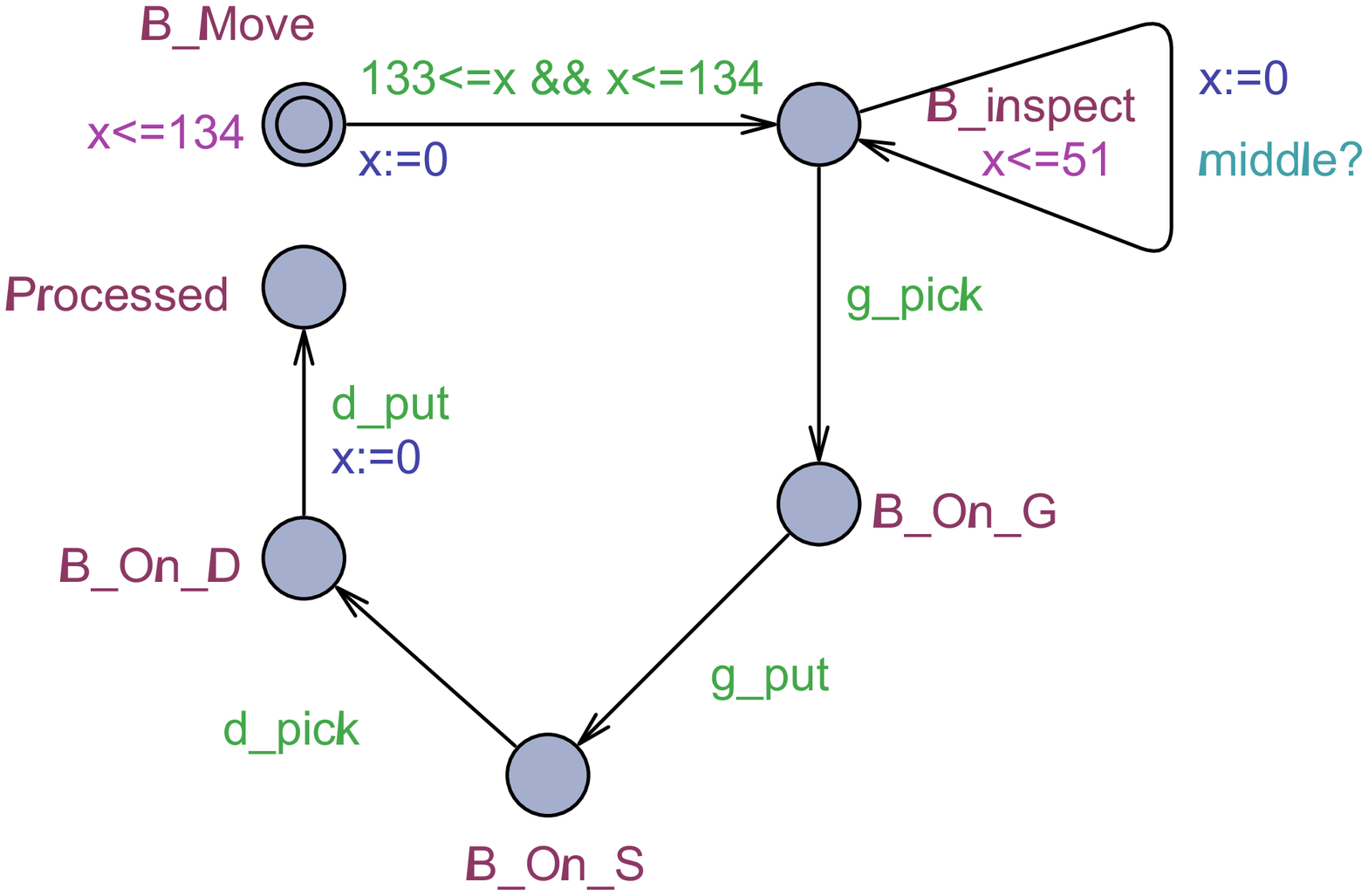}
    \caption{The Box template}
    \label{fig:Box}
  \end{minipage}
  \hspace{0.5cm}
  \begin{minipage}[b]{0.4\linewidth}
    \centering
     \includegraphics[width= 2.4in]{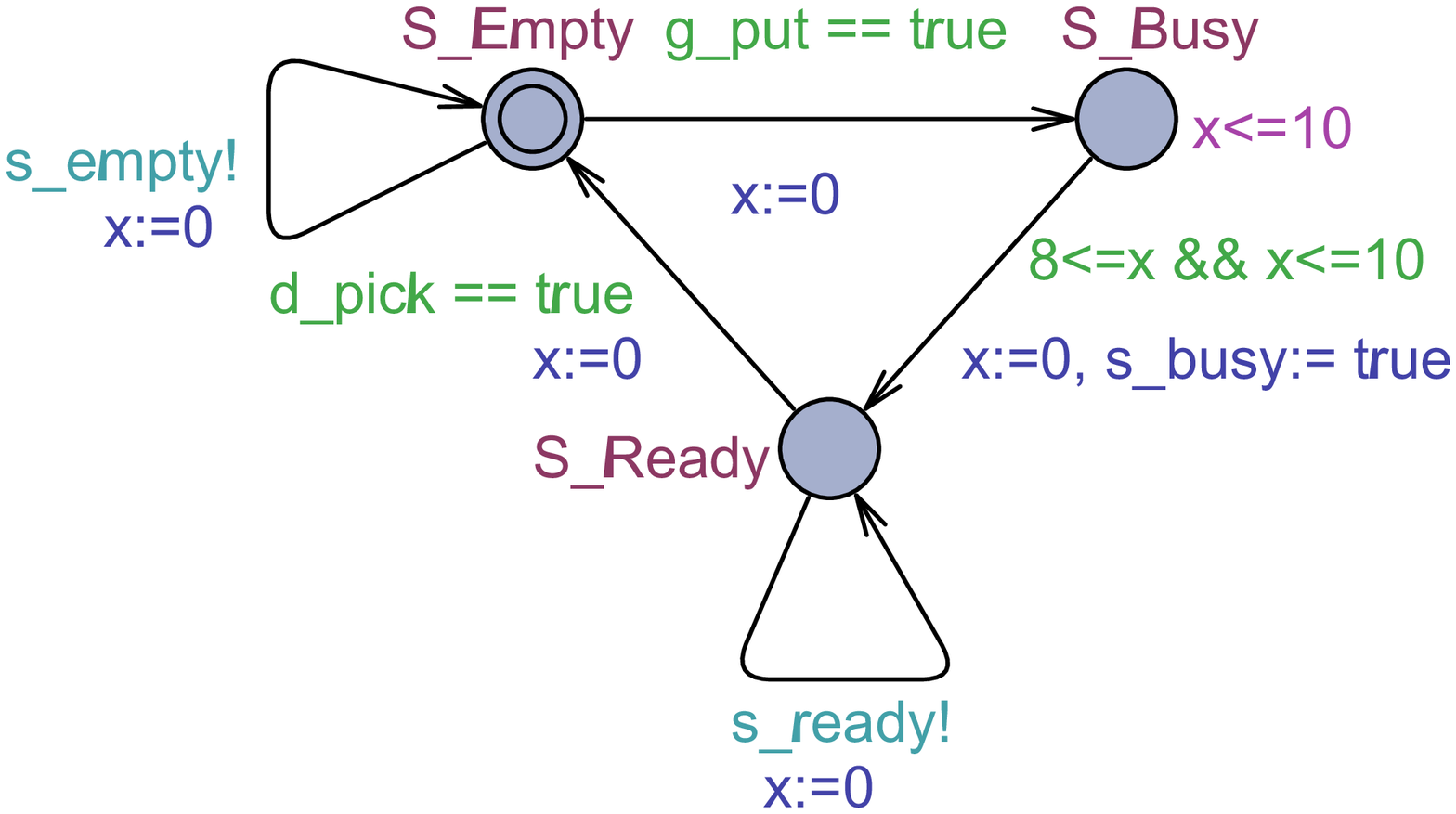}
    \caption{The processing station template}
    \label{fig:processing_station}
  \end{minipage}
\end{figure}

The timed automaton for the processing station is given in Figure \ref{fig:processing_station}.
The station is initially empty. Once a box arrives at the station it takes around 
8-10 time units to be processed. The box is then ready to be picked up by the D-Robot.

The timed automaton for the box is given in Figure \ref{fig:Box}.
The box initially moves from the left end of the belt to the inspection point.
It takes between 133-134 time units for the box to reach the inspection point
from the left end of the belt.
Then it will be picked up by the G-Robot.

\begin{table} 
  \small
  \centering
\begin{tabular}{|c|c|c|c|}
\hline
No. of processes & Run-time& Memory  & WCET \\
\hline 
  4&    0.015s &38,072KB &   171 \\
\hline
6  &  0.922s&45,860KB &    185 \\
\hline
9  &  72s&524,096KB&   215  \\
\hline
14  &   280s& 524,096KB&  325  \\
\hline
\end{tabular}  
\caption{The WCET of the manufacturing system for different number of boxes where the two robots move at different speeds \label{table:table1}}
\end{table} 

\begin{table} 
  \small
  \centering
\begin{tabular}{|c|c|c|c|}
\hline
No. of processes &  Run-time& Memory & WCET \\
\hline 
  4 &    0.015s&38,072KB &   153 \\
\hline
6  &  0.922s&45,860KB &    174 \\
\hline
9  &   70s&524,096KB&   204  \\
\hline
14 &   280s&524,096KB&  255  \\
\hline
\end{tabular}  
\caption{The WCET of the manufacturing system for different number of boxes where the two robots move at the same speed \label{table:table2}}
\end{table} 

Using the zone-based algorithms we could analyse the manufacturing system
up to 9 processes (automata) (6 boxes, G-Robot, D-Robot, and a service station).
 All experiments are conducted on a PC with 32-bit Redhat Linux 7.3 with Intel (R)
core CPU at 2.66 GHz and with 4 GB RAM.  
In Table 2 we verify the performance of the system under the following time constraints: 
the time required for the box to reach the inspection point is within $[133,134]$,
and the time required to process a box at the station is within $[8,10]$. In this configuration, we assume that the D-Robot
is faster than the G-Robot in the sense that the D-Robot can turn left and right and pick up and put boxes
faster than the G-Robot as shown in Figures \ref{fig:D-Robot} and \ref{fig:G-Robot}.
 
As we expect when we increase the number of boxes in the model the value of WCET varies which
implies that the number of boxes in the model impacts directly the WCET.
In Table 3 we verify the system under the same settings used in Table 2
except that we increase the speed of the two robots and assume that both robots move at the same speed.
In this configuration the time the robot takes to pick the box up or to put it down  is within $[1,2]$ time units, and the time it takes to turn left or right is within $[2,6]$ time units.
As shown in Table 3 the performance of the system under this configuration has been improved
where the values of WCET decreased under this configuration.

We also verified the WCET of the four TA given in Figures \ref{fig:infinite}, \ref{fig:simpleInfinite}, \ref{fig:infiniteSecond},  and \ref{fig:infiniteWithsThreeFinite}, which have an infinite WCET. The algorithm handles successfully these TA in a very reasonable time where each one of them has been verified in a few seconds. On the other hand, UPPAAL fails to terminate when verifying the automata in Figures \ref{fig:infinite}, \ref{fig:simpleInfinite}, \ref{fig:infiniteSecond}, and \ref{fig:infiniteWithsThreeFinite} and hence no answer has been obtained. In fact, we have verified in UPPAAL several other examples of TA with infinite cycles which show that when infinite cycles exist, the algorithm implemented in UPPAAL may not terminate.

\section{Conclusion and Future Work}
 
 In this paper we proposed an algorithm for determining the WCET 
in timed automata by modifying the underlying model-checking algorithm, 
rather than analysing those times by augmenting the models with clock variables and querying those.
The algorithm uses a modified abstraction which we call partial extrapolation that keeps the extra clock precise to the end of the analysis.
The proposed algorithm can work on any arbitrary TA including those containing infinite cycles.
For future work, we aim to develop some acceleration techniques for computing
WCET of subset of TA, namely those that contain paths with a large number of cycle iterations since our presented algorithm does not work very well
for such finite cycles. We believe that techniques based on some syntactical analysis for the behaviour of the cycle can be developed which may help 
to accelerate the WCET computations.

\bibliographystyle{alpha}
\bibliography{references}

\appendix

\section{Proofs}

\subsection{Proof of Lemma \ref{fixedpointTheorem}}

In order to prove this Lemma \ref{fixedpointTheorem} we need to show that after a finite number of iterations 
of $\pi$ while applying the $M$-extrapolation the search will reach a fixed point of the cycle.
Let $E_{\pi} = (e_0,.., e_{m-1})$ be the sequence of edges of the cycle $\pi$ 
and that $M(\mathcal{A})$ is the largest integer constant that appears in the guard and the location invariants of $\mathcal{A}$.
Since $\pi$ is an infinite cycle we can describe its behaviour using the recursive function $f_{\pi}(Z_s^{n}) = succ(Z_s^{n}, E_{\pi}) = Z_s^{n+1}$, where $Z_s^{n}$ is the corresponding zone at the start location of $\pi$ at iteration $n$.
We need to show that the zone approach with the $M$-extrapolation operator
guarantee the convergence of fixed point computations of $\pi$. That is, there will be two distinct iterations $j, k$ of $\pi$ such that $Z_s^{j}= Z_s^{k}$.
Before proceeding further in the proof, let us recall how a zone $Z$ is extrapolated when some clocks in $Z$ exceeds the bound $M(\mathcal{A})$. From the definition of $M$-extrapolation we know that when $D_{0,y} < -M(\mathcal{A})$ the extrapolation function sets $D_{0, y}$ to $-M(\mathcal{A})$ and when $D_{y, 0} > M(\mathcal{A})$ it sets $D_{y, 0}$ to $\infty$. Note that the domain of the lower bound of the clocks is still finite after extrapolation where for each clock $y \in X$ we have $(- M(\mathcal{A}) \leq D_{0, y} \leq 0)$.
The domain of the upper bound of the clocks is also finite after extrapolation.
It is necessary to note that $\infty$ is just a special value that we assign to a clock variable
when it exceeds the bound $M(\mathcal{A})$.
From the assumption that $\pi$ is infinite we know that there will be an infinite sequence of zones of the form $\verb+MExtra+_M(Z_s^{1}), \verb+MExtra+_M(Z_s^{2}),.., \verb+MExtra+_M(Z_s^{n}),..$ every time $\pi$ is executed.  Now given that the number of clocks in $\mathcal{A}$ is bounded ($|X| < \infty$)                                                                                                                                                                                           and the number of edges of $\pi$ is bounded ($ |E_{\pi}| < \infty$)                                                                                                                                                                                          and that each clock has a finite domain where for each $y \in X$ we have $(0 \leq D_{y, 0} \leq M(\mathcal{A}))$ or $ D_{y, 0} = \infty$ 
and $(- M(\mathcal{A}) \leq D_{0, y} \leq 0)$ and from the fact that $\pi$ is infinite in the sense that each location in $\pi$ will be visited infinitely often then it is easy to see that there will be
two distinct iterations $i$ and $j$ of $\pi$ where $ \verb+MExtra+_M(Z_s^j) = \verb+MExtra+_M(Z_s^k)$
and hence $ succ(Z_s^{j}, E_{\pi}) =  succ(Z_s^{k}, E_{\pi})$.

\subsection{Proof of Lemma \ref{optLemma1}}

Let first us denote  the matrix $\verb+MCanon+ (\verb+MExtra+_{m}(Z))$ as $Z^{'}$. Let us denote also the constraint $(c_{i, j}, \prec_{i, j})$ after extrapolating and canonicalizing it as $(c_{i, j}^{'}, \prec_{i, j}^{'})$. We need to show that $(c_{i, j}, \prec_{i, j}) = (c_{i, j}^{'}, \prec_{i, j}^{'})$. From the assumption that $Z$ is on canonical form and by definition of the tightening algorithm \cite{RokickiPhD} we know that the weight of the constraint $(c_{i, j}, \prec_{i, j})$ in $Z$ is the tightest weight that can be derived from the set of constraints in $Z$. By the definition of extrapolation we know that the bounds of the constraints that are extrapolated are in fact increased. Note that when the bound of a constraint is above the extrapolation constant $M$ then the extrapolation function $\verb+MExtra+_{M}(Z)$ sets it to $\infty$, which is an increase, and when the bound is less than $-M$ it sets it to $-M$, which is still an increase, and therefore the function $\verb+MExtra+_{M}(Z)$ only increases bounds. Now since the constraint $(c_{i, j}, \prec_{i, j})$ has not been increased during extrapolation and that $Z \subseteq Z^{'}$ and the function $\verb+MCanon+(Z)$ computes minimum it is easy to see then that $c_{i, j} = c^{'}_{i, j}$ and hence no need to recanonicalize $(c_{i, j}, \prec_{i, j})$.

\subsection{Proof of Theorem \ref{partial-extrapolation}}

Theorem \ref{partial-extrapolation} can be proved by reasoning about how $\land$, $\Uparrow$, \verb+reset+, \verb+MExtra+, and \verb+MCanon+ operations
together with the conditions (C1-C4) modify the zones of the resulting graph of an automaton.
The Theorem  can be proved by induction on the length of the transition sequences. Suppose we have an automaton $\mathcal{A} = (\Sigma, L, L_{0}, L_{F}, X,I, E)$ with a set of symbolic runs $\mathcal{R}$.
Let  $ r$  be an arbitrary run of $\mathcal{R}$ which can be either finite or infinite run. 
As induction hypothesis, assume that the entries $-D_{0, i}^k$ and $D_{i, 0}^k$  maintain respectively the precise infimum and supremum accumulated delays of the automaton $\mathcal{A}$ up to $k$-transitions (i.e. $\langle l_0, D_0 \rangle \leadsto^{\alpha,  k} \langle l_k, D_k \rangle$, where $\alpha$ is either a delay or discrete action). Assume further that $D^{k}_{i, j}$ and $D^{k}_{j, i}$,  where $j \neq i \land j =0,.., n$, maintain the precise upper and lower bound difference between the extra clock and each other automaton clock up to $k$-transitions. From the definition of the zone approach we can write the entries $D_{0, i}^k$ and $D_{i, 0}^k$ as follows
$D_{0, i} ^k = ( (D_{0, i}^{k-1} \land D_{0, i}(I(l_{k-1})))\Uparrow \land D_{0, i}(I(l_{k-1})) \land D_{0, i}(\psi_{k-1}))$
and $D_{i, 0} ^k = ( (D_{i, 0}^{k-1} \land D_{i, 0}(I(l_{k-1})))\Uparrow \land D_{i, 0}(I(l_{k-1})) \land D_{i, 0}(\psi_{k-1}))$.
From the semantic definition of DBMs and the fact that time can only elapse at locations
the two entries can be simplified as follows: $D_{0, i} ^k = x_0 - (\delta +\Sigma_{j=0} ^{k-1} (\inf  (d_j)))$ and $D_{i, 0} ^k = (\delta + \Sigma_{j=0} ^{k-1} (\sup  (d_j))) - x_0$,
where $d_j$ is the allowed delay interval at location $l_{j}$ such that for all $v \in d_j$ the invariant $I(l_j)$ holds.
Since $x_0$ has always the value 0 and $\delta$ has initially the value 0, 
we can then simplify the entries as follows: $D_{0, i} ^k = - \Sigma_{j=0} ^{k-1} (\inf  (d_j))$ and $D_{i, 0} ^k = \Sigma_{j=0} ^{k-1} (\sup  (d_j))$.
Now assume $\langle l_k, D_k \rangle \leadsto^{\alpha} \langle l_{k+1}, D_{k+1} \rangle$. 
We need to prove that after executing transition ($k+1$) the extra clock $\delta$ remains precise. 
That is, the entries $D_{0, i} ^{k+1} = D_{0, i} ^k  - \inf  (d_{k+1})$ and  $D_{i, 0} ^{k+1} = D_{i, 0} ^k + \sup  (d_{k+1})$, where $d_{k+1}$ is the allowed delay interval at location $l_{k+1}$.
As we know from the semantic definition of the zone approach the upper bound of the clocks may become (temporarily) imprecise at each step of the successor computation due to the application of the delay operation which sets the upper bound of all clocks to $\infty$ and hence the extra clock $\delta$ may become imprecise. However, during canonicalization and with the help of the diagonal  constraints of the form $(D_{i, j}, \prec_{i,j})$ and $(D_{j, i}, \prec_{j,i})$,  where $j \neq i \land j =0,.., n$, the exact upper bound of $\delta$ can be reconstructed. Hence, to show that the entries $(D_{i, 0}, \prec_{i,0})$ and $(D_{0, i}, \prec_{0,i})$ remain precise in the end we need to show also that all the diagonal constraints $(D_{i, j}, \prec_{i,j})$ and $(D_{j, i}, \prec_{j,i})$ remain precise during the analysis. However, since there are two types of transitions in TA: delay $\alpha = \epsilon(d)$ and action $\alpha \in \Sigma $ we need to consider two cases.
\begin{itemize}

\item (Delay $\alpha = \epsilon(d)$). By the assumption  $\langle l_k, D_k \rangle \leadsto^{\epsilon(d)} \langle l_{k}, D_{k} + \epsilon(d) \rangle$
we know that $D_{k} + \epsilon(d) \models I(l_k)$.
From the definition of $\leadsto$ and by delay we have $\langle l_k, D_k \rangle \leadsto \langle l_k, D_{i+1} \rangle$.
Expansion by the definition of $\Uparrow$ and $\land$ with the invariant at the location $l_k$ 
we get $D_{k+1} \in ((D_{k} \land I(l_k))^{\Uparrow} \land I(l_k))$.
By the definition of canonicalization and following condition C2 and C3 we get $D_{k+1} \in  (\verb+Mcanon+(D_{k} \land I(l_k))^{\Uparrow} \land I(l_k))$. 
From the semantic definition of DBMs and after executing the above operations we get $D_{0, i} ^{k+1} = D_{0, i} ^k  - \inf (d)$ and $D_{i, 0} ^{k+1} = D_{i, 0} ^k + \sup  (d)$. 
From the semantic definition of the zone approach we know that the first canonicalization operation will fix the non-tightness introduced by the delay operation and hence the exact lower and upper bound of $\delta$ will be reconstructed during this operation.
By induction hypothesis we know that $D_{0, i} ^k$ and $D_{i, 0} ^k$ maintain the accumulated delays of $\mathcal{A}$ up to $k$-transitions.
Thus the value of the entries $D_{0, i} ^{k+1}$ and $D_{i, 0} ^{k+1}$ in the matrix $(\verb+Mcanon+(D_{k} \land I(l_k))^{\Uparrow} \land I(l_k))$ are precise and hence the delay transition does not affect adversely the constraints involving $\delta$.

\item (Action  $\alpha \in \Sigma $). 
By the assumption $\langle l_k, D_k \rangle \leadsto^{\alpha} \langle l_{k+1}, \verb+reset+[\lambda] D_{k} \rangle$ 
we know $l_k  \xrightarrow{\psi, reset[\lambda]} l_{k+1}$.
From the definition of $\leadsto$ we have $\langle l_k, D_k \rangle \leadsto^{\alpha} \langle l_{k+1}, D_{k+1} \rangle$ 
by $l_k  \xrightarrow{\psi, reset[\lambda]} l_{k+1}$ if $D_{k+1} \in (D_{k} \land I(l_k))^{\Uparrow} \land I(l_k) \land \psi)$.
Expansion by the definition of $\land$ with the guard at the transition $l_{k+1}$  we get  $D_{k+1} \in \verb+MCanon+(\verb+MExtra+(\verb+MCanon+(\verb+MCanon+\\ ((D_{k}  \land I(l_k))^{\Uparrow} \land I(l_k) \land \psi)$. Expanding this by the \verb+reset+ operation
we get $D_{k+1} \in \verb+MCanon+(\verb+MExtra+(\verb+MCanon+(\verb+MCanon+((D_{k}  \land I(l_k))^{\Uparrow} \land I(l_k) \land \psi) [\lambda:=0])))$. 
Expand this by intersecting the resulting zone with the target invariant of location $k+1$ and extrapolate and canonicalize afterwards we get $D_{k+1} \in \verb+MCanon+(\verb+MExtra+(\verb+MCanon+(\verb+MExtra+(\\\verb+MCanon+(\verb+MCanon+((D_{k}   \land I(l_k))^{\Uparrow} \land I(l_k) \land \psi) [\lambda:=0])) \land I(l_{k+1}))))$. From the semantic definition of the zone approach we know that the second canonicalization operation  will fix the non-tightness introduced by intersecting the zone with the guard $\psi$ and will not affect adversely the constraints involving $\delta$. Thus the value of the entries $D_{0, i} ^{k+1}$ and $D_{i, 0} ^{k+1}$ in the matrix $(\verb+MCanon+(\verb+MCanon+((D_{k}  \land I(l_k))^{\Uparrow} \land I(l_k) \land \psi) [\lambda:=0])$ are still precise. From Lemma \ref{optLemma1} we know that all the constraints involving $\delta$ will not be changed  during the last canonicalization operation since the constraints involving $\delta$ will not be changed during extrapolation and that extrapolation only increases bounds while canonicalization computes a minimum. 
From the semantic definition of $\land$ and the \verb+reset+ operation
and that $\delta_i \not \in \lambda$ it is easy to see that the weight of the constraints involving $\delta$ in the matrix $\verb+MCanon+(\verb+MExtra+(\verb+MCanon+(\verb+MExtra+(\verb+MCanon+(\verb+MCanon+((D_{k}  \land I(l_k))^{\Uparrow} \land I(l_k) \land \psi) [\lambda:=0)) \land I(l_{k+1}))))$ are precise. Thus the value of the entries $D_{0, i} ^{k+1}$ and $D_{i, 0} ^{k+1}$ after executing the action transition ($k+1$) represent the precise infimum and supremum accumulated delays of $\mathcal{A}$ up to ($k+1$)-transitions.
\end{itemize}

It remains to show that the partial extrapolation approach ensures termination and correctness when the run $r$ is infinite or when there is an infinite cycle (i.e. cycle that can be repeated infinitely often) in $\mathcal{A}$. Note that the modified zone approach visits all the reachable states of an automaton while \textbf{ignoring the value of the extra clock $\delta$ (the non-extrapolated clock) for termination}. Hence, termination is guaranteed because there are finitely many sets of the form $\verb+MExtra+_{\mathcal{A}}(D)$. Also by condition C4 we know that cycles in TA will be treated differently during the analysis where the fixed point abstraction will be used rather than the inclusion abstraction when handling the generated zones inside cycles. This allows to detect whether the cycle can lead to an infinite WCET. Hence, the partial extrapolation approach ensures that the clock $\delta$ will be exact in the end even when there is a run with infinite length.

\subsection{Proof of Theorem \ref{mainThrm}}

Theorem \ref{mainThrm} can be proved by induction on the length of transition sequences.
However, the proof of the theorem is a straightforward combination 
of Theorems \ref{partial-extrapolation} and \ref{fixedpointTheorem} and corollary \ref{finiteCycleTheorem}.
From Theorem \ref{partial-extrapolation} (the partial extrapolation theorem) we know that the extra clock $\delta$ remains precise to the end and is not influenced by extrapolation. 
From Theorem \ref{fixedpointTheorem} we know that if there is an infinite cycle in the behaviour of $\mathcal{A}$ then it will be detected during the analysis since the algorithm checks at each iteration of the loop whether the search has reached a fixed-point of the discovered cycle. Now by checking the upper bound of the extra clock (the non-extrapolated clock) the algorithm can detect whether the reached fixed point is for an infinite cycle at which time can elaps or for an infinite cycle at which time cannot elapse. 
From corollary \ref{finiteCycleTheorem} we know that if there is a finite cycle in the behaviour of $\mathcal{A}$ then the cycle will be repeated the maximum allowed number of times and that the precise delays at each visited location are respected by partial extrapolation and hence it guarantees the precise calculations of WCET. Also from Theorems \ref{partial-extrapolation} and \ref{fixedpointTheorem} we know that termination is ensured since there are finitely many sets of the form $\verb+Extra+_{\mathcal{A}}(Z)$.

\end{document}